\documentclass[12pt]{article}
\usepackage{epsf}
\def\subsubsection{\@startsection{subsubsection}{3}{\z@}{-3.25ex plus
 -1ex minus -.2ex}{1.5ex plus .2ex}{\large\sc}}

\newcommand{\be}{\begin{equation}}
\newcommand{\ee}{\end{equation}}
\newcommand{\bel}[1]{\begin{equation}\label{#1}}
\newcommand{\bea}{\begin{eqnarray}}
\newcommand{\eea}{\end{eqnarray}}
\newcommand{\ba}{\begin{array}}
\newcommand{\ea}{\end{array}}

\newcommand{\ket}[1]{\mbox{$| \, {#1}\, \rangle$}}
\newcommand{\exval}[1]{\mbox{$\langle \, {#1}\, \rangle$}}

\def\bbbz{{\mathchoice {\hbox{$\sf\textstyle Z\kern-0.4em Z$}}
{\hbox{$\sf\textstyle Z\kern-0.4em Z$}}
{\hbox{$\sf\scriptstyle Z\kern-0.3em Z$}}
{\hbox{$\sf\scriptscriptstyle Z\kern-0.2em Z$}}}}

\begin{document}

\title{\Large Critical phenomena and universal dynamics
in one-dimensional driven diffusive systems with two species of particles}
\author{Gunter M. Sch\"{u}tz\\
\it \small $^\dagger$Institut f\"ur Festk\"orperforschung, Forschungszentrum 
J\"ulich,\\
\it \small D-52425 J\"ulich, Germany}
\date{\today}
\maketitle

\begin{abstract}
Recent work on stochastic interacting particle systems with two 
particle species (or single-species systems with kinematic constraints)
has demonstrated the existence of spontaneous symmetry breaking, long-range 
order and phase 
coexistence in nonequilibrium steady states, even if translational
invariance is not broken by defects or open boundaries. If both particle
species are conserved, the temporal behaviour 
is largely unexplored, but first results of current work on the 
transition from the microscopic to the macroscopic 
scale yield exact coupled nonlinear hydrodynamic equations and
indicate the emergence of novel types of shock waves which are collective 
excitations stabilized by the flow of microscopic fluctuations.
We review the basic stationary and dynamic properties of these systems,
highlighting the role of conservation laws and kinetic constraints for
the hydrodynamic behaviour, the microscopic origin of domain wall (shock) 
stability and the coarsening dynamics of domains during
phase separation.
\end{abstract}
\newpage
\section{Introduction}
\subsection{Why we are interested in driven diffusive systems}

The investigation of interacting particle systems far from equilibrium has
shown that one-dimensional driven diffusive systems with
short-range interactions exhibit a remarkably rich variety of
critical phenomena. Unlike in thermal equilibrium one observes 
spontaneous symmetry breaking, long-range order and phase coexistence in the 
steady state if the system evolves under certain microscopic kinetic 
constraints or has more than one conservation law.
A large body of work has been devoted to microscopic stochastic lattice
models for driven diffusive systems where classical interacting
particles move under the action of a random force preferentially
in one direction.\footnote{Strictly speaking
one should call these models mesoscopic as interactions on scales
below the particle size are replaced by effective
interactions. However, the notion microscopic
has become standard and will be used here. It is justified in relation
to a truly macroscopic description where particle positions are replaced by
coarse-grained density fields.}

The ongoing interest in these systems
has many reasons. The most obvious one is derived from a fundamental task of 
statistical mechanics, viz. the desire to understand the
emergence of macroscopic collective properties from microscopic
interactions, with a view on general features such as interaction
range (short- or long-ranged), kinetic constraints or the presence of 
conservation laws. It has turned out that parts of
that program can be carried out to a very satisfactory degree 
in the simplest case of driven diffusive systems of identical
conserved particles with hard-core interaction. 
Despite their simplicity, these systems
exhibit a rich and rather non-trivial dynamical and stationary
behaviour. For an exactly 
solvable paradigmatic model,
the asymmetric simple exclusion process (ASEP, see below)
not only the macroscopic nonlinear hydrodynamics have been derived rigorously
\cite{Kipn99,Ligg99} but also
detailed information about {\it universal} phenomena, including 
shock diffusion \cite{Ferr94b}, the microscopic origin of the stability of
shocks \cite{Schu00} and the dynamical structure 
function \cite{Prae02} could be obtained in the past decade.
It is then natural to ask what to expect in the presence of
more than one conservation law, i.e., in systems
with several distinct species of particles).

Secondly, in the absence of a general framework for studying
nonequilibrium systems (analogous to the usual principles of
equilibrium statistical mechanics), one needs to understand 
coarse-grained dynamical properties not only for their own sake, but also
in order to predict what stationary states these systems evolve into.
For one-species systems with open systems this has led to a theory of 
boundary-induced
phase transitions which provides a general framework for a quantitative
description of the steady-state selection in driven diffusive systems which
are in contact with particle reservoirs at their boundary. Unlike in 
equilibrium, boundary conditions determine the
bulk behaviour of driven diffusive systems in a decisive fashion
which can be captured in terms of an extremal principle for the
current \cite{Krug91a,Popk99}. The resulting phase diagram
for the nonequilibrium steady state
is determined by the interplay of localized excitations and
shocks. Again it is natural to ask for principles
of steady-state selection and the resulting phase diagram in 
systems with many species of particles which are characterized by
a conserved current for each particle species.

A third motivation for studying these systems stems from numerical
evidence which shows that addressing these questions not only leads 
to an encyclopedic accumulation of knowledge. Rather it was found that 
there is exciting new physics
in systems with more than one species of particles, including
spontaneous symmetry breaking and phase separation phenomena, even 
in translation invariant systems \cite{Evan00,Muka00} without the beneficial 
``assistance'' of open boundaries or static defects \cite{Jano97}
in facilitating phase transitions. It also emerged that 
similar phenomena may occur in 
single-species systems with one conservation law, provided there are
kinetic constraints determined by a zero-rate
condition somewhat analogous to the zero-temperature condition for
long-range order in equilibrium systems with short-range interactions.
Neither the hydrodynamic behaviour of systems with more than
one conservation law nor the microscopic conditions
for the occurrence of critical phenomena are well-understood.

These are some -- and by far not all -- fundamental reasons to investigate
translation invariant driven diffusive systems and to clarify the
role of conservation laws and kinetic constraints for their dynamical and 
steady state properties. Other important issues include the calculation of
exact stationary distributions and
large deviation functions, nonequilibrium Yang-Lee-theory and
phase transitions in systems with absorbing states. Some of these
topics are discussed in a complementary review by Evans \cite{Evan03}.
We also refer to \cite{Schm95,Priv97,Marr99,Hinr00} 
where closely related questions in nonconservative systems and higher 
dimensions are treated and to \cite{Ziel02} and references therein
for phase transitions in systems with continuous state space.
Here we focus on the issues of hydrodynamic behaviour, the microscopic origin 
of domain wall (shock) stability, spontaneous symmetry breaking and 
coarsening of domains in one-dimensional conservative lattice systems, with
particular emphasis on models with two conservation laws and on
exclusion processes with one conservation law in so far they are relevant in 
the context of the nonequilibrium bulk phase transitions that we review here.

\subsection{Basic models}

The asymmetric simple exclusion process \cite{Ligg99,Schu00}
has become a paradigmatic
example for a driven diffusive system and has begun to attain a status
in the study of nonequilibrium systems somewhat similar to the role
the Ising model plays in equilibrium statistical mechanics.
In this stochastic lattice gas model each lattice site is occupied by at most
one particle. Particles hop randomly in continuous time to the right 
neighboring site with 
rate $D_r$ and to the left with rate $D_\ell$ respectively, provided
the target site is empty. Otherwise the attempted move is rejected.
We present this hopping rule as follows:
\bea
\label{1-1}
A0\quad \to\quad 0A & \mbox{\rm with rate} & D_r \nonumber\\
0A\quad \to\quad A0 & \mbox{\rm with rate} & D_l .
\eea
Hopping attempts take place independently
with an exponential waiting time distribution with mean $D_r+D_\ell$ (Fig.~1). 

These simple rules specify completely the stochastic bulk dynamics
of the system.\footnote{For a mathematically precise definition of the process,
see \cite{Ligg99,Ligg85}.} For a finite lattice with $L$ sites one has to 
specify 
boundary conditions. Mostly commonly studied are periodic boundary conditions, 
reflecting boundaries (hopping confined to a box), and open boundary
conditions where particles may enter and exit the lattice at the boundary
sites $1$ and $L$ respectively with rates $\alpha,\beta,\gamma,\delta$
(see Fig. 1). By choosing $\alpha= \rho_1 D_r \lambda_1$,
$\gamma= (1-\rho_1) D_l \lambda_1$ as left boundary rates
and $\beta= (1-\rho_2) D_r \lambda_2$, $\delta= \rho_2 D_l \lambda_2$
the open system may be interpreted as being connected to
particle reservoirs with constant density $\rho_1$ at the left (auxiliary) 
boundary site $0$ and density $\rho_2$ at the right (auxiliary) 
boundary site $L+1$ respectively. The parameters $\lambda_{1,2}$ are
introduced to describe a hopping mechanism between the reservoirs
and the chain which may differ from the hopping inside the chain.
From a physics point of view this would correspond to
activation energies for entering (leaving) the system which are
different from those for hopping in the bulk.

This is the simplest model that incorporates the basic features of a driven
diffusive system with short range interactions. The short range interaction
is taken care of by the hard-core exclusion constraint. The randomness of the
hopping events models diffusive motion of the free particles outside the
interaction range. The hopping asymmetry corresponds to a
driving force that leads to a biased average motion and hence to a macroscopic
particle current even in the stationary state of the system, except in
the case of reflecting boundary conditions, where the system reaches
an equilibrium state \cite{Ligg85,Sand94a}.
Throughout this paper we assume a bias in positive lattice direction.

Notice that each lattice site
can be in two states, either occupied or empty, and hence the
system can be described in terms of occupation numbers $n_k=0,1$.
Implicit in this description
is the absence of any internal degree of freedom that particles may possess.
Hence all particles are indistinguishable. The number of particles is
conserved in  the bulk, but not at the boundaries in the case of the open
system. The single bulk conservation law gives rise to a current
via the lattice continuity equation
\bel{1-2}
\frac{d}{dt} \rho_k = j_{k-1} - j_k
\ee
for the expected density $\rho_k = \exval{n_k}$, averaged over realizations
of the stochastic time evolution and also averaged over different initial
distributions. For the ASEP the current follows straightforwardly
from the definition,
\bel{1-4}
j_k = D_r \exval{n_k(1-n_{k+1})} - D_\ell \exval{(1-n_k)n_{k+1}}.
\ee

For periodic boundary conditions there is a family of stationary distributions
which are Bernoulli product measures with density $\rho$, i.e.,
at each given site the probability of finding a particle is given by $\rho$,
independent of the occupation of other sites. According to
(\ref{1-4}) the stationary current
\bel{1-5}
j = (D_r - D_\ell) \rho(1-\rho)
\ee
is a nonlinear function of the density with a single maximum at
$\rho=1/2$. 

If the ASEP is confined to a box (corresponding to
``open boundaries'' with $\alpha=\beta=\gamma=\delta=0$) the system evolves
into an equilibrium state where essentially all $N$ particles form
a cluster of macroscopic size $\approx N$ with density $\rho\approx 1$ and 
the current vanishes. The density profile has a non-trivial
form (deviating significantly from 0 or 1 respectively) only at the
left edge of the cluster, denoted below as ``shock'', ``domain wall'' or 
``interface'' respectively, depending on context. The width of this
interface is finite on the lattice scale, i.e., microscopic. In the
grandcanonical ensemble there are no correlations, but the density profile 
has the form of a hyperbolic tangent \cite{Ligg85}. 
The canonical distribution is more complicated, with correlations
within the interface region, but has a similar density
profile \cite{Sand94a}. 

The open ASEP has a intriguing phase diagram with a nonequilibrium first order
transition at $\rho_1=1-\rho_2$ between a low-density phase with bulk density 
$\rho=\rho_1$ to a high-density phase with bulk density $\rho=\rho_2$. There are
nonequilibrium second-order transitions from both phases to a maximal
current phase with $\rho=1/2$, irrespective of the boundary densities
in the square defined by $\rho_1>1/2,\rho_2<1/2$. The density profiles
are non-trivial in all phases \cite{Schu93b,Derr93b}. At the first-order
transition line one has phase coexistence with domains of densities 
$\rho_{1,2}$, separated by a microscopically sharp domain wall (shock),
the position of which performs a random walk over the whole lattice. The 
exact solution of the stationary density profiles and a theory
of boundary-induced phase transitions which provides a microscopically
oriented derivation of the phase diagram is reviewed in detail in 
\cite{Schu00}. The bulk densities as a function of boundary densities
were obtained by Liggett \cite{Ligg75} using probabilistic methods.
More generally, the theory of boundary-induced phase transition which
takes into account the flow of fluctuations
has revealed that the bulk density that simple open driven diffusive systems
select can be obtained from the extremal principle \cite{Popk99}
\bea
 j & = &
\max_{\rho \in [\rho_{2},\rho_{2}]} j(\rho) \mbox{ for } \rho_{1}>\rho_{2} 
\nonumber \\
\label{min}
 j & = &
\min_{\rho \in [\rho_{1},\rho_{2}]} j(\rho) \mbox{ for } \rho_{1}<\rho_{2}
\eea
involving the stationary current-density relation that can be obtained
e.g. from measurements or exact calculations in periodic systems
where the density is conserved. The boundary densities $\rho_{1,2}$
entering (\ref{min}) are non-universal functions of the rates at which
particles enter and leave the system. For $\rho_2=0$ the extremal
principle was first proposed by Krug \cite{Krug91a} on a phenomenological
basis.

One may relax the exclusion constraint to allow for up to $m$
particles on each lattice site. This gives rise to the partial exclusion
process \cite{Schu94,Sepp99} with $m+1$ states per site. Also particle
systems with nonconserved internal degrees of freedom such as velocities
in traffic flow models \cite{Chow00,Helb01} have more than one possible
state per site, but still obey a single continuity equation of the form
(\ref{1-2}). The form (\ref{1-4}) of the current, however, strongly
depends on the microscopic hopping rules. An important model without
exclusion is obtained by the following simple mapping of the ASEP: Since the 
order
of particles is conserved in the ASEP one may regard
particles as sites of a new lattice gas system and the number of
vacancies $n_j$ between particles $j,j+1$ as occupation numbers on site $j$.
This gives rise to a special case of the zero-range process (ZRP) 
\cite{Spit70}. The particle hopping
rates in the ASEP turn into the rates of decreasing the number of particles
by one unit in the ZRP, with hopping to the right in the ASEP 
corresponding to hopping to
the left in ZRP and vice versa (Fig.~1).

The general homogeneous ZRP allows for hopping of a particle from site $j$ 
with a rate
$w_n$ that depends only on the unrestricted occupation number $n_j$.
Here we shall consider only nearest neighbour hopping (Fig.2) with
asymmetry factors $p,q$ to the right and left respectively.
Below a critical density (which may be infinite, see Sec.~4), the periodic
and infinite system has a family of stationary
distributions which are product measures where the probability of finding
$n$ particles on a given site is given by \cite{Spit70,Andj82}
\bel{1-6}
p_0 = \frac{1}{Z}, \quad p_n = \frac{1}{Z} z^n \prod_{i=1}^n 1/w_i
\ee
with the ``fugacity'' $z$ fixing the mean particle density and the
one-site nonequilibrium analog
\bel{1-7}
Z = \sum_{n=0}^\infty z^n \prod_{i=1}^n 1/w_i
\ee
of the partition function. The density as a function of $z$ is then given
by
\bel{1-8}
\rho = z \frac{d}{dz} \ln{Z}.
\ee
According to its definition via the
lattice continuity equation for the ZRP the stationary
particle current is given by 
\bel{1-9}
j = (p-q) \sum_{n=1}^\infty p_n w_n= (p-q) z.
\ee
The density dependence of $j$ in the stationary state can be obtained by
inverting the relation $\rho(z)$ which is a monotonically increasing
function of $\rho$ \cite{Evan00}. Notice that the radius of convergence
depends on the choice of rates $w_i$ (see Sec. 4).

In order to describe a system with two different
conserved species $A,B$ of identical
particles (or alternatively: tagged particles or particles with two internal
states which do not affect its dynamics) one needs a model where each lattice 
site can be found in at
least three different states: empty, or occupied by either an $A$-particle
or a $B$-particle. The most simple extension of the exclusion process
that accounts for the possibility of two particle species may hence
be described by the six hopping rates
\bea
\label{1-10}
A0\quad \to\quad 0A & {\rm with\quad rate } & D_{A0} \nonumber\\
0A\quad \to\quad A0 & {\rm with\quad rate } & D_{0A} \nonumber\\
B0\quad \to\quad 0B & {\rm with\quad rate } & D_{B0} \nonumber\\
0B\quad \to\quad B0 & {\rm with\quad rate } & D_{0B} \\
AB\quad \to\quad BA & {\rm with\quad rate } & D_{AB} \nonumber\\
BA\quad \to\quad AB & {\rm with\quad rate } & D_{BA}. \nonumber
\eea
There is no established name for this generic process and we shall refer to it
as two-species ASEP. Associated with the two conservation laws there are
two currents defined by
\bea
\label{1-11}
\frac{d}{dt} \rho^A_k = j^A_{k-1} - j^A_k \\
\frac{d}{dt} \rho^B_k = j^B_{k-1} - j^B_k .
\eea
Notice that in general $j^A$ and $j^B$ depend on both occupation numbers
$n^A_k, n^B_k$ respectively. Hence one has two coupled lattice continuity
equations. The stationary distribution of this process and hence
the current-density relation is known only
on certain parameter manifolds, see below.

The natural order parameter that describes the macroscopic state of the system
is the particle density of each species. Hence for each conservation law
there is an associated order parameter. Notice, however, that one may have
conservation laws that are only indirectly related to particles densities
which by themselves may not be conserved. E.g. in a reaction-diffusion
system $A+B\to 0$ \cite{Redn97}
where $A$ and $B$-particles annihilate upon encounter 
(i.e., react
into an inert reaction product), the difference $s=n^A - n^B$ still gives
rise to a single conservation law with an associated current, even though
$n^A$ and $n^B$ are not individually conserved. By interpreting
$A$-particles ($B$-particles) as carriers of a positive (negative)
electrical charge, one could speak in this case of charge conservation.
In the two-species catalytic reaction $A+B\to B+B$ \cite{Alim00}
the total density $n=n^A + n^B$
is conserved. Analogously, one may also consider models 
with two conservation laws, but more than three states per site. Examples
include e.g. two-lane models \cite{Lahi97,Popk01} or bricklayer models 
\cite{Toth02}.

There is no answer to the question to which extent or under which 
circumstances the existence of internal degrees of freedom matters for the
macroscopic properties of driven diffusive systems. Since, however, it is
now clear that the number of conservation laws is important, we shall
categorize the models in the following according to  this property.

\subsection{Closely related models, not covered in this review}

\noindent {\bf Equivalence to 2-dimensional equilibrium systems:}
The time-dependent one-dimensional stochastic processes discussed above
are equivalent to two-dimensional equilibrium systems, defined by some
vertex model \cite{Baxt82}. The time evolution is encoded in the transfer 
matrix of the 2-d model, e.g., the discrete-time ASEP with a
sublattice parallel update corresponds to the 
six-vertex
model \cite{Kand90,Schu93a,Hone96,Pigo02}. For three-states models the 
construction is entirely analogous
and leads to higher vertex models. Recently also integrable vertex models with
an unlimited number of states 
have been investigated \cite{Alca99a}. They
correspond to the zero-range representation of the ASEP.
The stationary
distribution of a one-dimensional process with $L$ sites
corresponds to the equilibrium 
state of the associated two-dimensional model, defined on a strip
of dimension $L\times \infty$. Processes defined in continuous time
are derived from the transfer matrix of the vertex model in the same
way as one obtains quantum spin-chain Hamiltonians \cite{Schu00}. Hence the
Markov generator of the stochastic time evolution is equivalent to
some (usually non-hermitian) one-dimensional quantum Hamiltonian.
This opens the tool box of condensed-matter physics for the study
of stochastic dynamics.
The ASEP (\ref{1-1}), the two-species ASEP (\ref{1-10}) on a certain parameter
manifold \cite{Popk02b} and also various single-species \cite{Alca94,Schu95a} 
and two-species reaction-diffusion models 
\cite{Alim00,Alca94,Dahm95,Fuji97,Mobi01} correspond to integrable models, for
which Bethe ansatz and related methods yield exact results on
the dynamics of the system.

We note that by considering the stationary states of one-dimensional
non-equilibrium systems as equilibrium distributions of a
two-dimensional systems the occurrence of long-range order and
phase separation becomes somewhat less mysterious. However, the important
question of how these phenomena emerge from the microscopic laws of
interaction cannot be answered by this formal equivalence. 
We mention in passing that in another mapping the 
steady-state distribution describes the equilibrium properties of a directed 
polymer in a 2-d random energy landscape \cite{Krug91b}.\\

\noindent {\bf Higher-dimensional nonequilibrium models:} 
The extension of the models
discussed above to higher dimensions is technically rather straightforward
and obviously of importance. For suitably chosen initial distribution some 
features of shocks may be present in higher-dimensional driven diffusive
systems, but many of the
physical properties of the one-dimensional systems
discussed below are expected to change dramatically not only because
of the upper critical dimension $d_c=2$ for diffusion (which makes
mean-field behaviour more likely to describe the systems), but also
because of the absence of blocking effects due to hard-core exclusion.
Moreover, there may be phase transitions in the unbiased equilibrium 
counterparts
of the model which lead to new phenomena in the biased case.
Two-dimensional driven diffusive systems are reviewed in detail
in \cite{Schm95}.\\

\noindent {\bf Many conservation laws:} A natural, but for the scope of
this review too far-reaching question is the behaviour of particle
systems with many conservation laws. Not surprisingly, no general picture
has emerged yet. An interesting generalization of the ASEP comprises
lattice systems with particles covering more than one lattice site,
but moving only by one site in each infinitesimal time step 
\cite{Sasa98,Alca99,Ferr02,Laka02,Shaw03}. The stationary distribution of 
these systems
has still a simple product structure and remarkably these models
are also integrable. This includes also polydisperse models with particles 
of different sizes (and hence as many conservation laws).

Another class of models with many conservation laws arises from
assigning to each particle
its own intrinsic hopping rate. In the zero-range mapping one thus
obtains a process with site-dependent quenched random hopping rates.
For rates drawn from some 
distribution the hydrodynamic behaviour has been studied in Refs. 
\cite{Benj96,Sepp99b,Barm02}. For certain distributions the system 
with asymmetric hopping rates undergoes at some critical density
a transition to a platoon state, where fast particles are trailing a
slow one reminiscent of traffic flow. This transition is a classical analog 
of Bose-Einstein
condensation \cite{Evan96,Krug96} and is quite analogous to the condensation 
transition in ordered systems to be discussed in detail below.
A similar model with passing of particles has also been studied 
\cite{Kari99a,Kari99b}. Remarkably, the phase diagram of the open
system both with \cite{Khor00} or without passing 
\cite{Beng99} has a structure similar to that of the 
usual one-species exclusion process \cite{Schu93b,Derr93b}
which can be explained by the theory of boundary-induced phase
transitions for systems with one conservation law \cite{Popk99,Kolo98}.

\section{Applications}
\setcounter{equation}{0}

The motivation we gave for studying the systems reviewed here addressed 
general
questions of non-equilibrium statistical mechanics, with little reference
to actual realizations where such processes might play a role. Applications
are actually numerous and include not only quasi one-dimensional settings
(e.g. molecular diffusion in nanoporous materials such as zeolites
\cite{Niva94,Kukl96,Lei96,Jobi97,Song00,Meer00}, single-file diffusion of 
mesoscopic 
colloidal particles \cite{Wei00}, or ionic conduction in narrow
channels \cite{Katz84,Sand95,Chou99}
but also -- through various mappings -- two and
three-dimensional systems. Of course, the basic models (\ref{1-1}),
(\ref{1-10}) can serve only as very crude approximations for any real
complex system. However, the universality of critical phenomena
(dynamical and static), of diffusion, of the
emergence of shocks and of coarsening allow for the study of fundamental 
properties
of real systems in terms of
simple toy models. It is not the purpose of this article to review
such applications in any detail, but some significant results are 
summarized.\\

\noindent {\bf Tracer diffusion:} The simplest way of obtaining a
system with two conservation laws consists in considering tagged
particles in the usual ASEP. Tagged particles (=particles of type B)
have the same physical properties as usual particles, except that they carry
a marker which allows for their identification, but does not affect
the dynamics. Thus one gets the two-species ASEP (\ref{1-5}) with 
\bea
\label{2-1}
D_{B0} & = & D_{A0} \nonumber\\
D_{0B} & = & D_{0A} \\
D_{AB} & = & D_{BA} \quad = 0. \nonumber
\eea

In the unbiased case $D_{A0} = D_{0A}$ a single tracer particle in a
stationary system of density $\rho$ is predicted to perform
anomalous diffusion with a mean square displacement 
$\exval{X^2(t)} \propto (1-\rho)/\rho \sqrt{t}$ \cite{Alex78a,vanB83,Arra83}. 
Recently this was confirmed experimentally in the investigation of tracer
diffusion in zeolites \cite{Kukl96} using pulsed field gradient NMR 
\cite{Karg92}
and in the study of single-file diffusion of colloidal particles \cite{Wei00}.

Also a driven tracer particle in an environment of unbiased $A$-particles
behaves subdiffusively with a square-root power law for the
mean square displacement \cite{Burl96}. In the fully driven case 
(\ref{2-1}) the situation is more complex. When averaging over random
initial states of the system according to the weights given by the
stationary distribution, the mean square displacement was proved to
grow linearly in time with a diffusion coefficient $D = (D_r - D_\ell)
(1-\rho)$ \cite{deMa85}. On the other hand, for fixed initial states
(averaging only over realizations of the process) the variance is expected
to grow subdiffusively with power $t^{2/3}$ \cite{vanB91,Maju91}.
In a finite system with periodic boundaries the variance in the number
of hops made in the totally asymmetric process ($D_{0A}=0$) has been
calculated exactly in the infinite-time limit \cite{Derr93a,Derr97a} 
and been found 
to decrease asymptotically $\propto 1/\sqrt{L}$ in system size. This is
to be expected from dynamical scaling with the well-known dynamical
exponent $z=3/2$ of the asymmetric exclusion process \cite{Gwa92,Kim95}.
\\

\noindent {\bf Shock tracking:} The ASEP exhibits shocks which on a
macroscopic level appear as stable moving discontinuities in the density
profile. It is of great interest to understand the microscopic structure
of shocks, i.e., the density profile and correlations on the microscopic
lattice scale which for a real system is the analogue of intermolecular
distances. The fundamental question is whether the density changes
quickly over molecular length scales or much slower (but still
abruptly on macroscopic scales). Trying to answer this question
leads to the problem of defining a microscopic
shock position in a given realization of the process. This can be
accomplished by introducing a second class particle, i.e., a particle
that moves w.r.t. vacancies like an ordinary (first class) particle, but 
behaves like a vacancy w.r.t. to ordinary particles
\cite{Ferr91}. This leads to rates
(\ref{1-10}) with 
\bea
\label{2-2}
D_{B0} & = & D_{A0} \nonumber\\
D_{0B} & = & D_{0A} \\
D_{AB} & = & D_{A0} \nonumber \\
D_{BA} & = & D_{0A}. \nonumber 
\eea
By studying the motion of a single second-class particle one finds
the mean shock velocity 
\bel{2-3}
v_s = \frac{j_1-j_2}{\rho_1-\rho_2}
\ee 
for a shock jumping from a density $\rho_1$
to $\rho_2$ with stationary current $j_{1,2}$ in each domain.
The expression (\ref{2-3}) may be deduced directly from mass conservation. 
For fixed
initial states the variance of the shock position is subdiffusive 
with power law $t^{1/3}$ \cite{vanB91}, while with averaging over
random initial states with stationary weights at different
densities $\rho_{1,2}$ to the right and left of the shock
(i.e., starting the system from a shock measure) one finds ordinary
diffusion. The diffusion coefficient was conjectured \cite{Spoh91} and 
subsequently proved \cite{Ferr94a} to be given by
\bel{2-4}
D_s = \frac{1}{2} \frac{j_1+j_2}{\rho_1-\rho_2}.
\ee
More detailed information about the microscopic structure of the
shock has been proved by a variety of methods, for a review see 
\cite{Ferr94b} and for more recent work \cite{Derr97,Ferr00,Beli02}
and references therein. It has been established that a shock is truely
microscopic in the sense that a rapid increase of the density of particles
occurs on the lattice scale. Loosely speaking one may say that the shock 
performs a random walk with drift velocity (\ref{2-3}) and diffusion 
coefficient (\ref{2-4}). Some details concerning the structure
of a shock are reviewed below.

In a region of smooth variation of the density the second class particle
allows for tracking localized perturbations of the density 
\cite{Prae02,Ferr93}. The mean velocity of the second class particle
is given by the collective velocity 
\bel{2-5}
v_c = \frac{\partial}{\partial \rho} j(\rho)
\ee
of a density perturbation. Its mean-square displacement grows superdiffusively
with power law 
\bel{2-6}
\exval{X^2(t)}-\exval{X(t)}^2 \propto t^{4/3}.
\ee
where the divergent effective diffusion coefficient $\Delta$
crosses over in a finite system to \cite{Derr99}
\be
\Delta \sim L^{1/2}
\ee
which is expected from dynamical scaling with dynamical 
exponent $z=3/2$.
As a toy model for econophysics the position of the second class particle 
marks the price
of an asset on the price axis in a limit order market, the 
first-class particles represent bid and 
ask prices \cite{Chal02}. The rigorous result (\ref{2-6}) corresponds to a 
Hurst exponent
$H=2/3$ for fluctuations which compares well with the empirically 
observed value $H\approx 0.6$ for intermediate time ranges. Introducing
annihilation and creation of particles (cancellation and renewal of orders)
leads to the Gaussian value $H=1/2$ after some crossover time, also in
agreement with empirical findings.\\

\noindent {\bf Traffic flow:} The occurrence and microscopic nature
of shocks in the ASEP
is reminiscent of traffic jams in vehicular traffic. Indeed, traffic flow
may be regarded as a driven diffusive system \cite{Chow00,Helb01}, 
albeit with a non-conserved
internal degree of freedom, viz. the speed of cars which is dynamically
determined by the competition of the desire to move at an optimal
high speed and the necessity to keep a velocity-dependent minimal
safety headway to the next
car. 
At low densities the mean distance between cars is larger than
the required safety headway and essentially all cars move at their optimal 
speed,
with some fluctuations. At high densities cars have a mean distance below
the safety headway corresponding to  the optimal speed which leads
to a slowing down of the traffic. As a result the mean current as
a function of density $\rho$
(known as fundamental diagram in the traffic literature) has a maximum
like the exclusion process, albeit with a much broader distribution of the 
current. In the first measurement of traffic flow in 1935 by Greenshields
\cite{Gree35} the measured mean flow of cars was approximated by the 
expression (\ref{1-5}) $j \propto \rho(1-\rho)$
that the ASEP yields. More sophisticated models that provide a much
better description of real traffic data have been developed in the
past decade, starting with the Nagel-Schreckenberg model \cite{Nage92}
which contains the ASEP as a simple limiting case. Yet some fundamental
features of the ASEP survive in the more complicated
Nagel-Schreckenberg model. The theory of boundary-induced phase transitions
developed in \cite{Popk99,Kolo98} which predicts the stationary phase
diagram of a single-species system with open boundaries 
in terms of the extremal principle (\ref{min}) explains
quantitatively
the phase diagram of the Nagel-Schreckenberg model in terms of
effective boundary densities and is also
consistent with measurements of real highway traffic \cite{Popk01b}.

A description of traffic flow with a lattice model with one conservation
law corresponds to modelling cars which all have the same intrinsic
optimal speed -- a rather crude approximation if describing mixed traffic of 
cars and trucks is envisaged. This naturally leads as a next approximation
step to a two-species description with ``fast'' and ``slow'' particles
respectively. A model with two conservation laws arises also from the
study of two-way traffic flow with interaction between lanes, but
no exchange of particles in each lane. Taking the exclusion process
as the simplest possible model one arrives at the model (\ref{1-10})
with suitably chosen rates \cite{Mall96,Lee97,Jafa00}, see below.
A mixture of cars with individual intrinsic speeds leads to the
disordered hopping models mentioned above.\\

\noindent {\bf Biophysics:} The ASEP with open boundaries
was first developed as a simple
model for describing the kinetics of protein synthesis \cite{MacD68,MacD69}.
Here particles are ribosomes moving along the codons of a messenger RNA.
Each codon corresponds to a specific aminoacid which the ribosome uses to
assemble a protein. When such a step is completed the ribosome moves on
to the next codon and continues with the addition of next aminoacid
to the growing protein molecule. The injection of particles at one end marks 
the initialization of the process, the absorption at the other boundary
describes the release of the ribosome. The shock known from the exclusion 
process
corresponds to a traffic jam of ribosomes which explains an experimentally
observed slowing down of the ribosomes as they approach the terminal point
of the m-RNA where they are released after completion of the protein
synthesis \cite{Laka02,Shaw03,Heij87,Schu97}. 

Very recent work has shown that
in another biological setting exclusion particles may describe molecular
motors such as kinesins moving along microtubuli or actin filaments
in a cell \cite{Lipo,Kolo}. Due attachment and detachment during the motion
a description with non-conservative dynamics where particles are annihilated
and created also in the bulk with a small rate is required. This leads to
the model of Ref. \cite{Chal02} with open boundaries which yields
interesting new phenomena \cite{Parm03}. Oppositely moving molecular
motors give rise to a two-species exclusion model, in analogy to 
two-way traffic flow with interaction between lanes. 

A two-species exclusion process has also been introduced to describe the 
motion of ants along ant trails \cite{Chow02}. While crawling along a trail, 
the ants -- modelled by $A$-particles hopping along a lattice
-- produce pheromones ($B$-particles) which serve as a marker of the 
traversed path for other 
ants which again produce pheromones for subsequent ants. This is necessary to 
stabilize the trail as the pheromones evaporate after some time. 
The pheromones
are modelled as an immobile particle species which is deposited when
a hopping event has taken place and which disappears with some  
evaporation rate. The analogy of the
flow of ants to traffic flow has been pointed out in Ref. \cite{Burd02} who
measured the flow rate versus the density of ants, i.e., the current
density relation. The numerical results obtained from the 
two-species ant trail model yield qualitatively similar results
\cite{Chow02}. Essentially the same model (with different parameter values
and update rules)
has been introduced as a ``bus route'' model where one observes bunching
of particles (=``buses'') as they travel along lattice (``bus stops'')
and pick up passengers \cite{Oloa98}. Bunching of real buses appears to
occur on services which do not run according to fixed schedules, but which
stop according to demand.\\

\noindent {\bf Polymer dynamics:} 
In polymer networks such as rubber gum or gels, in polymer melts or in
dense solutions of macromolecules such as DNA different polymer strands
form a complicated topological structure of entanglements somewhat
reminiscent of a large portion of spaghetti. The entanglements severely 
restrict the dynamical degrees of freedom of the polymer chains. In the 
framework of the
celebrated reptation theory \cite{MacL02} developed by Doi, Edwards, and de 
Gennes \cite{Doi86,deGe79} the motion of an individual polymer is viewed
as being confined by a hypothetical tube which models the collective effect
of all entanglements of the neighbouring polymer chains. In an uncrosslinked
melt or solution the tube is open at both ends, since at the end points the
motion of polymer segments transverse to its own contour is not restricted
by topological constraints. This picture results in a snakelike one-dimensional
effective dynamics of polymer segments along the tube, with extra
orientational degrees of freedom only at its ends. 

In a lattice model of Rubinstein \cite{Rubi87} the reptation dynamics is
modeled by the symmetric exclusion process (\ref{1-1}) with open boundaries
which describe the extra end point degrees of freedom. With this model
exact results for the
relaxation of the contour and contour length fluctuations have been 
obtained.
Recent experiments on the dynamics of single entangled DNA-molecules
in dense solution confirm the findings \cite{Perk94,Schu99}.

Duke \cite{Duke89} extended the model to allow for tracking the spatial
orientation of the tube rather than only its length. This was done
in order to introduce a reference axis for describing gel electrophoresis, 
i.e., the separation of polymer fragments according to their length $L$. By 
applying an electric field of strength $E$ (the direction of which is the 
reference axis) 
a charged polymer is expected to move through a gel matrix (which provides
an entanglement network) according to the rules of reptation. However,
standard reptation theory does not allow for a prediction of the drift 
velocity $v$ beyond the linear response regime of small fields or very long 
polymers where
\bel{2-7}
v \propto D E L.
\ee
Here $D$ is the diffusion coefficient of an entangled polymer, predicted
by reptation theory to scale
\bel{2-8}
D\propto 1/L^2
\ee 
with length. The extended Rubinstein-Duke model is an asymmetric
three-states exclusion process (\ref{1-10}) with $D_{AB}=D_{BA}=0$ and open 
boundaries. Exact and rigorous results \cite{vanL92,Prae96} confirm
the predictions (\ref{2-7}), (\ref{2-8}). Moreover, simulations at
high fields yield the drift velocity in the non-linear regime \cite{Bark94}
which are in good agreement with experimental data \cite{Bark96}.
At sufficiently high fields the model exhibits spontaneous symmetry
breaking in the orientation of the polymer chain \cite{Aalb96}.
The asymptotic behaviour (\ref{2-8}) of the diffusion coefficient
has also been proved to remain valid in the presence of quenched
kinematic disorder which is described a generalized
Rubinstein model with many conservation laws where particles have their 
individual hopping rates \cite{Will02}.

A long-standing mystery in reptation theory has been the asymptotic 
behaviour of the viscosity $\eta$ of a polymer melt which is expected to
scale asymptotically \cite{Doi86,deGe79}
\bel{2-9}
\eta \propto L^3.
\ee However, experiments consistently give higher value $\approx
3.4$ of the scaling exponent. Doi had suggested this to be a finite-size
effect due to tube-length fluctuations \cite{Doi86}. That tube-length
fluctuations lead to an increased effective exponent could be confirmed 
by a careful numerical analysis of the Rubinstein-Duke model \cite{Carl01}. 
Also details of the
end-segment dynamics were shown to have significant non-universal
impact on finite-size behaviour of the viscosity and the diffusion
coefficient \cite{Paes02}.\\

\noindent {\bf Spin relaxation:} By interpreting occupation numbers as 
classical spin variables 
$s_k = 1-2n_k$ the ASEP describes biased Kawasaki spin-exchange
dynamics \cite{Kawa66} for the one-dimensional Ising model at infinite 
temperature. At finite temperature, biased Kawasaki dynamics correspond to an 
exclusion process 
\bea
\label{2-9a}
XA0Y\quad \to\quad X0AY & {\rm with\quad rate } & D^{XY}_r \nonumber\\
X0AY\quad \to\quad XA0Y & {\rm with\quad rate } & D^{XY}_l .
\eea
with next-nearest neighbour interaction depending on the
occupation $X,Y$. With some constraints on the rates \cite{Katz84}
this model has a stationary distribution 
\bel{2-9b}
P(\underline{n}) = \frac{1}{Z}\mbox{e}^{-\beta (E(\underline{n})+\mu N)} 
\ee
which is an Ising measure
with energy $E=- J\sum_k n_kn_{k+1}$ and a magnetic field $\mu$ which plays 
the role of a chemical potential
in the lattice gas interpretation of the variables.
At sufficiently low temperatures
the particle (=spin) current becomes a non-convex function
of the density (=magnetization) and unusual phenomena such as splitting
of shock fronts which separate regions of different density
can be observed \cite{Popk99,Font03}. In spin language the three-states
model (\ref{1-10}) describes dynamics of a classical spin-1 system.\\

\noindent {\bf Interface growth:} It was already realized in the 1980ies
that the ASEP describes the dynamics of a fluctuating interface by
considering the spin variables as local discrete slopes of an interface
on a two-dimensional substrate \cite{Meak86,Plis87} (Fig. 2). Hopping of a 
particle
to the right between sites $k,k+1$ corresponds to the random deposition
of a particle on site $k$ the dual growth lattice, hopping to the left to an 
evaporation (Fig.~2). One thus obtains
a growth model in the universality class of the one-dimensional 
KPZ equation \cite{Kard86},
reviewed in \cite{Krug91b}. We stress that the mapping is not one-to-one.
Since in the exclusion presentation only the local slopes enter, the 
information about the actual height of the interface gets lost. One can
keep track of the height by introducing an extra random variable for the
local height at some reference point $k_0$, which is increased (decreased)
by 2 units whenever a particle hops across the bond $k_0,k_0+1$ to the
right (left). The steady state current of the ASEP then gives the average
growth velocity, while fluctuations of the current measure fluctuations
in the local interface height. The extension of this mapping to 
the generalized exclusion process (\ref{1-10}) is obvious, one obtains a system
where local height differences may take values $0,\pm 1$. Some growth
dynamics considered below have the property that particles cannot be chipped
off a complete layer. This corresponds to a hidden conservation law, which
cannot be expressed in terms of particle occupation numbers alone.

\section{Steady states and hydrodynamic limit}
\setcounter{equation}{0}

\subsection{Steady states for driven diffusive systems}

As has become clear above, the stationary behaviour, i.e., the state the
system evolves into, is the first question to be addressed in the investigation
of driven diffusive systems.\footnote{This is formally analogous to 
investigating
the equilibrium behaviour of a many-body system and hence in the mathematical 
literature stationary states are often referred to as equilibrium states, even
though the presence of macroscopic currents prevents the applicability of
the usual notions of equilibrium statistical mechanics such as reversibility
and detailed balance. We note, however, that as long as only a stationary 
distribution is concerned -- without reference to the stochastic dynamics
for which the distribution is stationary -- it is convenient to use notions
borrowed from equilibrium statistical physics such as partition
function or canonical/grandcanonical ensembles respectively.} 
In this article we are concerned with the behaviour of translational
invariant systems, defined either on a finite lattice with periodic
boundary conditions or on the infinite integer lattice $\bbbz$.
Consequently we shall investigate stationary distributions which are
either translation invariant or where translation invariance is
spontaneously broken. 

It is important to bear in mind
that entirely different dynamics may have the same stationary distribution.
Indeed, for any given distribution
one may always construct some equilibrium dynamics (obeying reversibility)
using the principle of detailed balance.
Moreover, a strongly nonequilibrium system may have the stationary
distribution of some equilibrium model, an example being the KLS model
(\ref{2-9a}).  Therefore, equality of stationary 
ensembles for different systems has no implications whatsoever on the 
dynamical properties of these models. We also remark that simple dynamical 
rules
may result in stationary distributions with a complicated structure
and long-range correlations (see below), while complicated dynamical
rules may very well lead to simple stationary distributions.
Some stationary distributions for one-species models have been reviewed
in the introduction, here we focus on the two-species ASEP.

Fortunately, not only the single-species
ASEP but also many two-species stochastic particle
systems of interest have simple stationary distributions, the simplest
being product measures with stationary probabilities of the
form
\bel{3-3}
P \propto \mbox{e}^{\mu^A N^A +\mu^B N^B}.
\ee
Here $N^{A,B} = \sum_k n_k^{A,B}$ are the conserved total particle
numbers of each species (for one-component systems one has $n_k^B=0$)
and $\mu^{A,B}$ are the corresponding chemical potentials. For fixed
$N^{A},N^{B}$ all configurations are equally likely.
In equilibrium, such measures correspond to non-interacting systems.

By writing the master equation for the stochastic dynamics in the quantum 
Hamiltonian formalism 
\cite{Schu00} it is
straightforward to determine what dynamics have stationary
product distributions. In this formalism a product measure is represented
by a tensor product vector $\ket{P}=\ket{p}^{\otimes L}$. Each
factor $\ket{p}$ in the tensor product has as components the probabilities
of finding a lattice in a given state.
Hopping events between sites $k,k+1$ are generated by
a local stochastic matrix $h_k$ acting nontrivially only on the terms
$k,k+1$ in the tensor product. The full time evolution is generated by the
stochastic Hamiltonian $H=\sum_k h_k$ and the
stationarity condition for $\ket{P}$ reads
\bel{3-4}
H \ket{P} =0.
\ee
Because of translational invariance a stationary product measure therefore
satisfies the relation
\bel{3-5}
h_k \ket{P} = (d_{k+1} - d_{k}) \ket{P}
\ee
with an arbitrary matrix $d_{k}$ acting nontrivially only
on site $k$. This relation is usually
very easy to verify. A similar approach can be chosen for more complicated
measures, e.g., Ising measures with stationary probabilities of the
form
\bel{3-6}
P \propto \mbox{e}^{-\beta\sum_k (J^{AA} n^A_kn^A_{k+1}+J^{AB} n^A_kn^B_{k+1}
+J^{BB} n^B_kn^B_{k+1})+\mu^A N^A +\mu^B N^B}.
\ee
Given these measures the current can be calculated exactly as a
function of the densities $\rho^A$, $\rho^B$ via the invertible relationship
between the chemical potentials and the densities. For the 2-states
ASEP (\ref{1-10}) one has a product measure on the parameter manifold
defined by
\bel{3-7}
D_{0A}-D_{A0} + D_{B0}-D_{0B} + D_{AB}-D_{BA}= 0
\ee
For a $K$-species ASEP with rates $D_{XY}$ for the hopping process
$XY \to YX$ there are $(K-1)(K-2)/2$ conditions for the existence of a 
product measure \cite{Arnd98a}
\bel{3-8}
D_{0X}-D_{X0} + D_{Y0}-D_{0Y} + D_{XY}-D_{YX}= 0.
\ee

In the quantum Hamiltonian formalism the stationary distribution is the
ground state vector of the associated quantum spin chain.
For integrable models \cite{Alca93,Popk02b}  one may use the Bethe-ansatz
and symmetry properties \cite{Sand94a} for the explicit construction
of stationary states which are not simple product measures. Using
the Bethe ansatz has not been attempted yet for this class of models. 
A more popular method is the application of the matrix
product ansatz, reviewed in \cite{Derr98}. In this approach one defines a 
product measure with matrix entries $D_m$ rather than $c$-numbers as 
stationary weights for finding a given site in state $m$.  
The matrices $D_m$ to together with a set of auxiliary matrices
\cite{Stin95a,Stin95b} have to satisfy algebraic relations
which are obtained from requiring the matrix product state to satisfy
the stationarity condition (\ref{3-4}). This leads to algebras with
quadratic relations 
\cite{Derr93b,Derr93a,Arnd98a,Kreb97,Hinr96c,Essl96,Mall97,Isae02}.

The matrix product construction is equivalent to writing
the stationary distribution of the lattice gas in terms of
a transfer matrix $C=\sum_{m=0}^K D_m$ of some $n$-states
equilibrium system, determined by the
representation of the algebra, in particular its dimension $n$
which may be finite or infinite.
It is clear that finite-dimensional
representations correspond to stationary states with exponentially
decaying correlations unless the largest eigenvalue of $C$ is degenerate.
Usual product measures (complete absence of correlations) correspond
to one-dimensional representations. For a detailed review, see 
\cite{Evan03,Derr98}. The approach can also be extended to describe the full 
time evolution and hence yield time-dependent
probabilities of the system evolving from some
nonstationary initial distribution \cite{Stin95a,Stin95b,Schu98b}.
Popkov et al. have identified the parameter manifold for which the
dynamics of the 2-species ASEP can be solved using the dynamical matrix
product ansatz \cite{Popk02b}. Such models are all integrable in the
sense of being associated with an integrable vertex model. The more
general models of Ref. \cite{Arnd98a,Isae02} for which only the stationary 
distribution can be constructed with matrix products include also 
non-integrable models.

\subsection{Steady states with one $B$-particle}

A series of intriguing results have been obtained for two-species
systems (\ref{1-10}) with
just one particle of type $B$. Conditioning on having a second class
particle at some given site and calculating the probability of finding
a first-class particle at distance $r$ yields the density profile
as seen from the shock position, defined by the position of the 
second-class particle. The density approaches
its asymptotic shock densities $\rho_{1,2}$ at an exponential rate,
given in a non-trivial way by the hopping asymmetry $D_r/D_l$
and the densities $\rho_{1,2}$ \cite{Derr97,Derr93c}. For a special value 
of the asymmetry
one has Bernoulli measures with densities $\rho_{1,2}$ to the left
and right respectively. This is the result of a $q$-deformed 
$SU(2)$-symmetry of the Heisenberg quantum Hamiltonian that generates 
the time evolution of the process. For this value of the asymmetry
(or, equivalently, arbitrary asymmetry, but special density $\rho_2$)
the time evolution of the shock measure has been calculated exactly
both for the continuous-time ASEP \cite{Beli02} and for a discrete-time
variant \cite{Pigo02}. The shock position performs a lattice random walk
with rates given by the currents and densities in the two branches
of the shock. Two consecutive shocks with densities $\rho_{1,2,3}$
which can be defined by two
second-class particles form a usual bound state with finite mean
distance and exponentially decaying distance distribution if
a condition on $\rho_{2,3}$ originating in the $q$-deformed 
$SU(2)$-symmetry is met \cite{Kreb03}. Generically
two second-class particles form a weak bound state with infinite mean
distance and algebraically decaying  probability $p(r) \propto r^{-3/2}$
of being a distance $r$ apart \cite{Derr97,Derr93c}.

Using the algebra arising from the stationary matrix product
ansatz Mallick \cite{Mall96}
has studied the two-species ASEP (\ref{1-10}) with rates 
\bel{3-9} 
D_{A0}=1,\quad D_{B0}=\alpha, \quad D_{AB}=\beta.
\ee
All other rates are zero. For $\alpha=1$, $\beta=0$ the ``impurity''
particle $B$ corresponds to a tracer particle, for $\alpha=\beta=1$
it is a second-class particle. This model describes``cars'' ($A$-particles)
and ``trucks'' ($B$-particles), with a passing rate $\beta$. For
$\alpha<1$ and $\beta<\alpha$ a single truck acts like an impurity, 
hindering the motion of cars.
The current of $A$-particles, the
velocity of the impurity and the density profile as seen from the
impurity have been calculated exactly \cite{Mall96}. The system with a 
single impurity
exhibits an interesting phase diagram as a function of the hopping rates.
In one of the phases the system develops a stationary shock for sufficiently large
density $\rho>\rho_c$, analogous
to a traffic jam building up behind a slow vehicle. One has coexistence
of a low-density domain and a high-density domain, separated by a
domain wall (Fig.~3). The diffusion constant of the impurity has also been
calculated exactly \cite{Derr99,Bout02}.

Lee et al. \cite{Lee97} considered the model (\ref{1-10}) with rates
\bel{3-10} 
D_{A0}=1,\quad D_{0B}=\gamma, \quad D_{AB}=1/\beta
\ee
corresponding to oppositely moving particles (slow-moving ``trucks''
for $\gamma<1$) which interact upon encounter. In the presence
of a single truck (impurity) the average speed of cars (and hence the current),
the speed of the truck and the density profile of cars has been
calculated using again the same stationary three-species algebra \cite{Lee97}.
One obtains a phase diagram with a transition to a jammed phase
at a critical density $\rho_c=1/\beta$ (Fig.~3), with the
remarkable property of having the same statistical properties as
a deterministic ASEP with a fixed impurity \cite{Schu93a}. In a finite
system of site $L$
the position of the (microscopically sharp) domain wall fluctuates over 
a region of length $\propto \sqrt{L}$.
In a system
with two trucks they form a weak bound state in the traffic jam phase.
Notice that the model (\ref{3-10}) is equivalent to (\ref{3-9}) by
exchanging $B\leftrightarrow 0$. However, in this mapping a single
truck corresponds to a single vacancy, a scenario not studied in
\cite{Mall96}.

Arndt et al. \cite{Arnd98b} introduced a model of type (\ref{3-10})
with 
\bel{3-11} 
D_{A0}=\lambda, \quad D_{0B}=\lambda, \quad D_{AB}=q,\quad D_{BA}=1
\ee
For finite densities of both particle species
this model is reviewed below. Jafarpour \cite{Jafa00}
considered the presence of a single $B$-particle and calculated exactly
for $\lambda=1$ the speed of $A$-particles and the impurity and the density 
profile
of $A$ particles. As in the model of Lee et al. there is a phase transition 
from a free-flowing to a jammed phase, here at a critical density 
$\rho_c=q/2$ (Fig.~3). 

We remark that the jamming transition seen in these
three versions of the two-species ASEP may be regarded as a kind of
condensation transition where a finite fraction of $A$-particles 
condenses into a macroscopic block trailing the moving impurity.
A similar transition occurs also in the usual ASEP with a fixed
blockage \cite{Schu93a,Wolf90,Jano92}, corresponding to an immobile 
$B$-particle and passing event $AB0 \to 0BA$ rather than $AB \to BA$.
Hence single particles in a system with two conservation laws play
a role somewhat similar to local inhomogeneities in a system with
one conservation law. Notice that
for a fixed blockage and for the model of Mallick the jammed phase
exists between two critical densities $\rho_c^-,\rho_c^+$.
Lee et al. and Jafarpour resp. report only a lower critical density,
below which the system is in the free-flow phase.

\subsection{Hydrodynamic limit for finite densities}

The link between the stationary state and the dynamics is established by the
continuity equation which relates the change in the local order parameters
to the currents and other equations for the dynamics
of the various non-conserved internal degrees of freedom. In order to
obtain information about the dynamics on a course-grained scale one
assumes local stationarity and a sufficiently smooth behaviour
of the local order parameter. For driven diffusive systems one may then
investigate the dynamics on the Euler scale, i.e., in the scaling
limit where the lattice spacing $a$ and time scale $\tau$ are sent to 0 
such that $a /\tau$ remains constant.

By expressing the correlation functions that enter the
currents completely in terms of the density one thus obtains a partial
differential equation
\be
\label{3-1}
\frac{\partial}{\partial t} \rho =  - \frac{\partial}{\partial x} j(\rho)
\ee
for systems with one conservation law, and system of equations
\bea
\label{3-2}
\frac{\partial}{\partial t} \rho^A & = &
- \frac{\partial}{\partial x} j^A(\rho^A,\rho^B) \\
\frac{\partial}{\partial t} \rho^B & = &
- \frac{\partial}{\partial x} j^B(\rho^A,\rho^B) 
\eea
for models with two conservation laws. Using the theory of partial 
differential equations this yields coarse-grained information about the
time-evolution of the lattice model. The average occupation of the
local density is thus described in terms of a deterministic evolution
of a coarse-grained density profile. 

From these introductory remarks it has become clear that knowing 
the stationary currents {\it exactly} as functions of the density is crucial
for calculating the density profile. A mean-field approximation which
neglects correlations between different lattice sites is bound to give 
wrong quantitative results unless the stationary distribution 
happens to be characterized by  the absence of correlations. For systems
with a single conservation law mean field
approximations may yield qualitatively correct behaviour if the mean-field
current reproduces local extrema and inflection points of the true 
current.
However, we wish to stress that even short-ranged correlations resulting from 
Ising-type stationary
distributions (\ref{2-9b})
may lead to qualitatively wrong behaviour of the current-density
relation. An example is the KLS-model (\ref{2-9a})
where the exact current has a local minimum between two symmetric maxima
\cite{Popk99,Bran91} whereas the mean-field approximation yields the
current-density relation (\ref{1-5}) of the ASEP
with a single maximum.
Hence neglecting short-ranged correlations may yield not only
quantitatively but even qualitatively
wrong predictions for the most basic dynamical properties of the system,
viz. the stationary current and the coarse-grained time evolution of the
density profile. To conclude, one needs to know which generic features
of a current density relation determine the qualitative behaviour of the
solution of the hydrodynamic equation and it must be verified that a
mean-field approximation reproduces those features. Otherwise a mean field 
treatment of the continuity equation yields no information about the dynamics 
of the system. In some cases approximating an unknown measure by an
Ising measure (\ref{2-9b}), (\ref{3-6}) with short-ranged
correlations rather than by a simple product measure (simple mean field)
without correlations may bring improvement. In the literature this 
improved kind of approximation scheme is sometimes called cluster 
approximation. In the following we assume that at least the stationary current
(if not the full measure) is known exactly.

Nonlinear equations of the form (\ref{3-1}), (\ref{3-2}) are known to possess
singularities which do not allow for a unique solution of the initial
value problem. Almost all initial configurations will develop
shock discontinuities where the density jumps from one value to
another even if the initial state was smooth. This raises the question 
of the microscopic
properties of the macroscopic solution. Moreover, ambiguities exist
even if the original particle problem has a unique
stationary state into which the system evolves for all initial states.
Thus one is faced with the second problem of selecting the physical
solution of the hydrodynamic equation (Fig.~4). 

For a class of systems with one conservation law
the transition from the stochastic lattice dynamics to the
hydrodynamic equation (\ref{3-1}) is mathematically well-understood 
\cite{Reza91} and includes also the treatment of shock discontinuities, 
for a broader overview see \cite{Kipn99}. Here we give a more physics oriented 
account inspired by the desire to derive macroscopic phenomena such as
shocks from microscopic behaviour, viz. the flow of localized 
(microscopically) perturbations inside a stationary region. It has turned 
out that such an approach, originally developed for systems
with one conserved density and no internal degrees of freedom,
also works for two-species problems, if suitably generalized. 

To keep the discussion simple we restrict the review to a system
with a convex current-density relation such as found in the ASEP (\ref{1-5}).
In the hydrodynamic limit (\ref{3-1}) one obtains for the ASEP the well-known
inviscid Burgers equation \cite{Burg74}
\bel{3-12}
\frac{\partial}{\partial t} \rho =  -\frac{\partial}{\partial x} 
(D_r-D_l)\rho(1 - \rho)=-(D_r-D_l)(1 - 2 \rho)\frac{\partial}{\partial x}\rho.
\ee
It is well-known that an upward shock in the ASEP constitutes a stable 
shock whereas the downward shock dissolves into a rarefaction wave.
This result can be obtained by using the method of characteristics which trace
the motion in space-time of points of constant density. One
introduces a scaling variable $u=x/t$ to obtain from (\ref{3-12})
\bel{3-13}
u = (D_r-D_l)(1 - 2 \rho).
\ee
This is a weak solution for an initial profile with a down-shock at $x=0$ (Fig.~4).

This solution may be obtained in a different way by starting from the
lattice continuity equation (\ref{1-2}) with current 
\bel{3-14}
j_k = (D_r-D_l)\rho_k(1-\rho_{k+1}).
\ee
obtained from (\ref{1-4}) by neglecting correlations at all
times. The picture underlying this approximation is the
assumption of local stationary where all correlations are sufficiently
small.
With a Taylor expansion in the lattice constant $a\to 0$
one arrives at the Burgers equation
\bel{3-15}
\frac{\partial}{\partial t} \rho =  -\frac{\partial}{\partial x} 
(D_r-D_l)\rho(1 - \rho) + \nu \frac{\partial^2}{\partial x^2} \rho
\ee
with an infinitesimal viscosity $\nu \propto a$. The Burgers equation
is integrable using the Hopf-Cole transformation 
$\rho = \kappa \partial_x \ln{w}$. This leads to a standard linear 
diffusion equation for $w$ which has a unique solution for any initial profile.
Taking the limit $a\to 0$ in the solution one recovers the physical
solutions described above (shock and rarefaction wave)
which are realized by the ASEP. 

In a different, but for one-species models
equivalent approach for selecting the physical solution one defines an
entropy function associated with the conservation law (\ref{3-1}).
The entropy solution yields the physical solution corresponding to the
ASEP, for a review, see \cite{Kipn99}. With this approach one may also
consider the ASEP with a localized defect \cite{Baha98} which has a 
different solution for a downward shock and which also produces a shock
inside a domain with constant density \cite{Schu96}.

In order to obtain a physical microscopic picture of how these 
solutions emerge on a macroscopic scale we study 
the dynamics of localized perturbations in a homogeneous stationary
environment \cite{Schu00}. The time evolution of such a perturbation on the 
lattice
scale
can be probed by examining the dynamical structure 
function 
\bel{3-16}
S(k,t) = \exval{n_k(t)n_0(0)} - \rho^2
\ee
which measures the density relaxation of a local 
perturbation in the stationary state of uniform density 
$\rho$. Generally, the width of such a perturbation at $t=0$ is of the order
of the bulk correlation length.
The center-of-mass velocity of the perturbation is given by
the collective velocity
\bel{3-17}
v_c = \frac{\partial}{\partial \rho} j(\rho)
\ee
One may derive this relation from the shock velocity
\bel{3-18}
v_s = \frac{j_1-j_2}{ \rho_1-\rho_2} 
\ee
by taking the limit $\rho_1\to \rho_2$ of the asymptotic
densities of the shock. Notice that $v_c$ changes sign at
local extrema of the current density relation. We stress
that the expression (\ref{3-18}) follows from mass conservation
and hence no specific assumptions on the nature of the microscopic
dynamics are involved. Hence also (\ref{3-17}) is expected to be
generally valid. The only assumption are sufficiently rapidly
decaying bulk correlations in the steady state as otherwise 
the microscopic definition of a shock position becomes questionable,
since the shock position cannot be defined on a scale below the
bulk correlation length.
A direct derivation of (\ref{3-17}) from the dynamical structure
function which uses only translational invariance, the conservation
law and decay of correlations is outlined in \cite{Schu00}.

{}The shock velocity for the ASEP follows from (\ref{1-5}), (\ref{3-18})
and one finds
\bel{3-19}
v_s = (D_r-D_l)(1 - \rho_1 - \rho_2).
\ee
This yields the collective velocity
\bel{3-20}
v_c = (D_r-D_l)(1 - 2 \rho).
\ee
The origin of the physical solution of the macroscopic time
evolution can now be explained  from a microscopic viewpoint by
imagining that by a small
fluctuation a certain amount of mass detaches from the shock and forms
a perturbation at a small distance from the shock position. On average
this fluctuation will then travel with speed (\ref{3-20}) where $\rho$
is to be taken as either $\rho_1$ or $\rho_2$, depending on whether the
fluctuation had originally moved to the left or right of the shock. Eq. 
(\ref{3-20}) shows in the case of an upward shock that for all 
shock densities $\rho_{1,2}$ 
\bel{3-21}
v_c^{(1)} > v_s > v_c^{(2)}.
\ee
Hence in the moving reference frame of the shock the excess mass drifts 
back to the position of the shock and hence stabilizes it. 

On the other hand,
in the case of an initial downward shock the fluctuating
excess mass moves on average away from the
shock. Therefore this shock is not stable against fluctuations, in
the course of time the shock smears out and develops into a rarefaction
wave. In order to predict the macroscopic shape of the rarefaction wave, 
we assume
an initial configuration 
with a shock with densities $\rho_1$ and $\rho_2$ 
which is composed of many infinitesimal subsequent shocks 
at various levels of intermediate densities (Fig. 3).
Neither of these shocks is stable, but each slowly dissolving shock
at density $\rho$ moves with a speed $v_c$. Hence we conclude
that on the Euler scale (where the spread of a perturbation (\ref{2-6})
and hence of the increasing width of the unstable shock is scaled to zero)
points of constant density $\rho$ generally move with speed $v_c$.
From this observation the explicit form of the rarefaction wave
can be constructed.

Clearly, this is not a rigorous argument.
Support for this picture comes from the hydrodynamic limit. The collective
velocity is then nothing but the speed of the characteristics of the 
corresponding hydrodynamical equation $\partial_t \rho = -\partial_x j$
resulting from the continuum limit of the lattice
continuity equation (\ref{1-2}).
In this limit, the criterion (\ref{3-21}) becomes the defining property
of a shock discontinuity \cite{Lax73}. It asserts that the characteristics 
are moving into the shock. Otherwise, the characteristics
yield the rarefaction wave, as rationalized above.
For current-density relations which are not
globally convex one decomposes a single shock into subsequent small
shocks and then applies (\ref{3-21}) to these minishocks in order to
decide on stability. By taking the limit of infinitesimal 
shocks one
recovers in this way the scaling solution of the hydrodynamical equation
obtained from the method of characteristics \cite{Popk99,Hage01}.

For systems (\ref{3-2}) with two or more conservation laws there is no 
well-established mathematical theory for the selection of the physical
solution in the corresponding lattice gas. In recent work \cite{Toth02}
T\'oth and Valk\'o have  obtained rigorous results  by making use of Yau's 
relative 
entropy method
\cite{Yau91} which essentially proves that a product measure with
time-dependent local densities evolving according to the solution of the
hydrodynamic equation converges to the true measure in the sense
of relative entropy of the two measures. This approach works for systems
with a stationary product measure until a shock has formed, provided some 
particular identities hold which relate the macroscopic fluxes in the 
hydrodynamic pde. These identities are reminiscent of the Onsager's
reciprocity relations. The systems studied in   \cite{Toth02}
are models with generically more states than conservation laws.
They include a family of 2-species ASEP's with parameters satisfying the 
relation (\ref{3-7})
guaranteeing the existence of a stationary product measure and
which have a natural interpretation as growth models.
A two-species zero-range process has been studied by Grosskinsky
and the corresponding hydrodynamic equations have been established
\cite{Gros03}. 

The rigorous approach of  \cite{Toth02} is rather powerful
but in its current state fails as soon as shocks develop.
On the other hand, shock waves and special rarefaction waves
have been analysed nonrigorously from a microscopic viewpoint by studying 
the flow of perturbations and 
correspondingly extending the physical arguments presented above
to 2-species systems \cite{Popk03}. One has to study
two perturbations in each conserved density
which due to the interaction are forced to move with the
same velocity. Generalizing the analysis of the dynamic structure 
function to two conservation laws one finds that 
the main difference to the case of one conservation
law is the evolution of two distinct pairs of perturbations out of a 
single pair. Each pair
moves with collective velocities $v_c^\pm$ given by the eigenvalues 
of the Jacobian
\bel{3-22}
{\cal D} = \left( \ba{cc} 
\frac{\partial}{\partial \rho^A} j^A & \frac{\partial}{\partial \rho^B} j^A\\
\frac{\partial}{\partial \rho^A} j^B & \frac{\partial}{\partial \rho^B} j^B
\ea \right).
\ee

Corresponding to the two pairs of perturbations a single shock splits into 
two separate shocks, leaving the
system in stationary regimes of three distinct densities, viz. the
left and right shock densities $\rho^{A,B}_{1,2}$ enforced 
by the initial
state and selforganized intermediate shock densities $\tilde{\rho}^{A,B}$.
The equality of each pair of shock velocities (given by the general
expression (\ref{3-18}) applied to each single shock) and the requirement
that the velocity of the left shock $v_s^L$ must be smaller than 
the velocity $v_s^R$ of the
right shock determines the 
intermediate densities $\tilde{\rho}^{A,B}$. 
Requiring that all perturbations be absorbed in the shock
 one arrives at the condition
\bel{3-23}
v_c^\pm(\rho_1) > v_s^L > v_c^-(\tilde{\rho}), \quad 
v_c^+(\tilde{\rho}) > v_s^R > v_c^\pm(\rho_2)
\ee
for shock stability in driven diffusive systems with two conservation
laws. Violating one of these conditions leads to rarefaction waves
which have partially been described \cite{Popk03}. The general features
discussed here are confirmed by Monte-Carlo simulation of a two-lane
model with a conserved density on each lane related to the models of
Refs. \cite{Popk01,Lahi97}, but with periodic boundary conditions
and different choice of hopping rates respectively.
A complete description of the evolution for all possible initial states
has not yet been achieved. 


\section{Critical Phenomena}
\setcounter{equation}{0}

It is well-known that in thermal equilibrium one-dimensional systems
with finite local state space and short range interactions do not 
exhibit phase transitions at positive
temperatures, only at $T=0$ long range order may exist. From a dynamical
viewpoint there are no thermal fluctuations at $T=0$ in a classical system. 
In a terms of a stochastic process that
means that all transition rates are zero. Conversely, if a transition
rate is non-zero, some dynamics -- not necessarily satisfying detailed
balance -- is going on and it has been conjectured
that quite generally a system with strictly positive transitions rates and 
local interactions can have at most one stationary distribution, which
is often rephrased by saying that there can be no phase transition in a 
one-dimensional system with strictly positive rates. One has in mind
an infinite system since in a finite system
dynamics with strictly positive rates
are always ergodic and the conjecture is trivially true.

To rationalize the
conjecture one imagines, in the simplest case, two potentially stationary
distributions characterized by a different value of the order parameter.
An example is the Ising model where the order parameter is the magnetization,
which can take two different values below the critical temperature
in two or higher dimensions. The reasoning behind the positive rates
conjecture is the difficulty to imagine a local mechanism that eliminates
islands of the minority phase (created constantly by thermal fluctuations
in a region where the other phase
dominates) since in one dimension energetic
effects due to line tension play no role. A local mechanism cannot
detect the size of a minority island, therefore
such an island can grow indefinitely
and destroy the majority phase. Since noise (implied by strictly
positive rates) can always create such islands there seems to be no
possibility to keep the majority phase stable against fluctuations.
In a certain ``natural'' class of systems with nearest-neighbour interaction
this conjecture has been proved rigorously some time ago \cite{Gray82}.

Therefore it came as a surprise that P. Gacs constructed a model on the
infinite lattice which
violates the positive rates conjecture \cite{Gacs86,Gacs01}.
However, both the model and the proof that there is a phase transition
is rather complicated \cite{Gray01}, requiring either a very large
local state space or a very large interaction range, and the quest for
simple models with this property continues to stimulate research. 
As a guideline we note that the conjecture is clearly true for dynamics 
satisfying detailed balance with respect to a local interaction energy
as in this case the stationary distribution is just
the usual equilibrium distribution and the argument underlying the
positive rates conjecture applies. Hence one should look for models
that either violate detailed balance or have a nonlocal interaction
energy, but local dynamics.

We address the question of phase transitions in a broader sense 
by asking whether
phenomena associated with phase transitions such as divergent length
scales or spontaneous symmetry breaking may occur far from equilibrium.
Divergence of some correlation length does not necessarily require
the existence of more than one stationary distribution for a given set
of system parameters. One the other hand, one could have a parameter
range with two or more
stationary distributions which are not related by any symmetry, but the 
transition
into this regime would be associated by a divergent length scale.

On the other hand we restrict ourselves to driven diffusive systems with one 
or two conservation laws, thus skirting the issue of phase transitions
in non-conservative models, addressed in 
\cite{Parm03,Evan02,Popk03b}, and also
avoiding systems kept out of equilibrium without a having current in the
conserved density. This could be achieved e.g. by coupling 
{\it symmetric} hopping
dynamics to heat baths of different temperatures, a scenario not envisaged
in the context of driven diffusive systems. Also non-conservative processes 
with absorbing states
such as the contact process \cite{Marr99,Ligg99,Hinr00} fall in the class of
systems not considered here.  We refer to models of this
type only where it serves to illuminate the properties of closely related
driven systems.

By definition, conservative systems have
a continuum of stationary states (characterized by the value of the
order parameter) and hence the critical phenomena we review 
concern transitions between different stationary distribution at the
same value of the order parameter and coexistence of macroscopic stationary
domains where the order parameter takes different values. The domain
walls separating these domains are the shocks discussed in the previous
section. Hence the stability of domain walls is intimately connected
with the existence of phase separation.

\subsection{One conservation law}

In a system with a conserved density the positive rates conjecture
does not apply by definition, as transitions violating density
conservation have zero rate. Hence there seems to be no reason to pursue
the question of existence of phase transitions in conservative systems. 
Moreover, the discussion of shock
stability presented above shows that a stable domain wall separating
regions of different value of the order parameter may exist, thus
actually suggesting the possibility of phase coexistence and long-range
order. However,
by the same reasoning it is clear that in a system with generic
current density relation one cannot have two domain walls which would
be necessary to have macroscopic phase separation in a translation
invariant system: If say, an upward shock from density $\rho_1$ to
density $\rho_2$ at the left boundary of the region of higher density
$\rho_2$ is stable by the criterion (\ref{3-21}), then the downward 
shock $\rho_2$ to $\rho_1$ at the other
boundary of the region of higher density would be unstable by virtue
of the same criterion. Thus one is forced to conclude that phase
coexistence in a system with one conservation law may exist only
in the absence of translation invariance. This is indeed known
in systems with open boundaries where external particle reservoirs 
enforce regions of constant boundary densities $\rho_{1,2}$, separated
by a single stable domain wall \cite{Derr93b,Schu93b,Kolo98}.
Similarly in a periodic, but not translation-invariant system with
a defect one may have phase separation since in this set-up the defect
may stabilize the intrinsically unstable shock.
Hence, stability of a shock in a system with one
conservation law does {\it not} constitute a violation of the zero-rate
conjecture for a translation invariant system. Indeed, it has been suggested
that one-dimensional driven diffusive systems do not exhibit long-range order
in their steady states \cite{Spoh91}.

Yet, several translation invariant models with one conservation law
and short-range interactions which exhibit a robust phase 
transition were discovered. A simple, but non-generic example is a growth
model (which can be mapped to driven diffusive system according to
the strategy explained in Sec. 2) where a roughening transition
from a phase with a smooth interface to a phase with a rough interface
occurs \cite{Savi85,Krug90}. This class of models is non-generic in so far
as there is an intrinsic maximal growth velocity of the interface, enforced 
by a discrete-time parallel update. The smooth phase and hence the
phase transition disappears if the limit of continuous time is taken
in these models. 

Addressing the possibility of a roughening transition
in systems with continuous time evolution Alon et al \cite{Alon96,Alon98}
proposed a two-species ASEP with rates
\bel{4-1}
D_{A0}=D_{0B}=(1-q)/2,\quad D_{0A}=D_{B0}=D_{BA}=q
\ee
and annihilation/creation rates
\bel{4-2}
D_{00} = D_{BA} = q, \quad D_{AB} = 1-q
\ee
for the transitions $00 \leftrightarrow AB$ and
$BA \to 00$ respectively. These dynamics lead to a single
conserved ``density'' $S=N^A-N^B$. In the mapping to a growth
model $A$ ($B$) represent a local slope 1 ($-1$) and $0$ represents slope 0.
In order to ensure periodic boundary conditions also in the interface
representation the model has been studied for $S=0$.
For small $q<q_c$ there is a smooth phase where a local mechanism
eliminates islands in a flat region, since islands are formed 
with
boundaries that are biased to move towards each other. This mechanism
applies for islands of all sizes (except completed layers) and hence
leads to a smooth interface, consisting mainly of vacancies in the
lattice gas picture. Above $q_c\approx 0.189$ the creation of new islands
overcompensates the disappearance due to their intrinsic tendency to shrink 
and the system is in a growing
rough phase with a finite fraction of particles. The growth model
is in the universality class of the KPZ equation which is represented
by the standard ASEP. The critical behaviour at $q_c$ which is
related to directed percolation is discussed in detail
in \cite{Alon96,Alon98} where also a version of the model without
constraint on the local slope (corresponding to the absence of exclusion
in lattice gas language) is discussed. There is spontaneous symmetry breaking
in the smooth phase which can be quantified by introducing either a
colouring scheme, giving vacancy clusters between $A,B$ pairs a colour,
or by introducing a nonconserved order parameter $M$ that makes use of the 
interface representation. The critical exponent $\theta$ associated with the
vanishing of the order parameter by approaching the critical point from
below,
\bel{4-3}
\exval{|M|} \sim (q_c-q)^\theta,
\ee
is a new exponent, not yet understood in the framework of
directed percolation.

For this transition to exist it is crucial that no chipping of particles
from the
interface in a locally flat environment may occur. This is ensured by
imposing a kinetic constraint on the local dynamics, viz.
setting the rate $\tilde{D}_{00}$ for the process $00 \to BA$ to zero. 
However, this
process does not violate the single conservation law. Hence this model
does not provide a counterexample against the zero-rate conjecture,
as applied to systems with one conservation law.
By regarding the various local transitions of this strongly
nonequilibrium process as induced by thermal
activation from heat bathes at different temperatures one is led to
conclude that
the local mechanism that guarantees bounded growth of regions of the minority
phase is bought with a zero-temperature condition on the chipping process.

In the presence of chipping with a rate $\tilde{D}_{00}=p$ the interface 
attains a negative 
stationary growth velocity for some value of $q$ that depends on $p$.
An interesting phenomenon then occurs if the dynamics of the interface
is constrained by the minimal height condition that $h_i \geq 0$ for all
times and all lattice points $i$ \cite{Hinr97}. For negative growth velocity 
the
interface is driven towards the hard wall located at the height level $h_i=0$
which the interface cannot penetrate.
In the special case of $p=1-q>1/2$ the model satisfies detailed
balance with respect to the energy 
\be
E=\sum_{i=1}^L h_i
\ee 
which is the area under the interface. The stationary probability of finding 
an interface configuration $\underline{h} = (h_1,\dots ,h_L)$ is given by 
\bel{4-4}
P(\underline{h}) = (q/(1-q))^{E(\underline{h})}/Z_L
\ee 
with the partition function $Z_L = \sum_{config} (q/(1-q))^E$.
At $q=1/2$ the interface has mean velocity
zero, for $q>1/2$ the interface grows. Hence the expression (\ref{4-4})
diverges in time and becomes meaningless as stationary distribution
for $q\geq 1/2$. In the growth regime the interface roughens, with a dynamical
behaviour in the ubiquitous KPZ universality class.
For $q< 1/2$ the interface is bound to the hard wall
and hence is smooth. Close to the transition point exact analysis of the 
partition function yields an occupation density $\sigma$ of the bottom layer 
$h=0$ and a width $w$ diverging as  \cite{Hinr97}
\bel{4-5}
\sigma \sim (q_c-q)^1, \quad w\sim (q_c-q)^{1/3}
\ee
The unbinding of the interface at $q_c=1/2$ is analogous to a wetting
transition. We remark that this model has a zero-rate constraint
by not allowing the interface to penetrate the bottom layer. 

The physics described by these two models can be 
captured in a generalized KLS-model (\ref{2-9a}). In this model
spontaneous symmetry breaking due to absence of chipping, and
conservation of minimal height with the
resulting wetting transition can be studied without reference to the
height variable.

Koduvely and Dhar considered the symmetric KLS-model with rates
$D^{XY}:=D^{XY}_{r}=D^{XY}_{l}$ for the hopping event
$XA0Y\leftrightarrow X0AY$ \cite{Kodu98}. The analog
of the chipping rate in the two-species model is the hopping rate 
$D^{A0}$. Setting $D^{A0}=0$ automatically leads to
conservation of minimal height, or, more precisely, conservation of the
height level of a completed layer. It is not necessary to stop the
dynamics by a separate rule involving the local height. 
In contrast to the previous model of Hinrichsen et al, however,
the interface always remains anchored
to the minimal height level at some random position. Hence the properties
of the wetting (unbinding) transition are described by this model only below
the critical point in the dry (bound) state. 

Careful numerical analysis of the symmetric model \cite{Kodu98}
indicates subdiffusive critical dynamics
of the unbiased interface with a dynamical exponent $z\approx 2.5$,
as opposed to the usual dynamical exponent $z=2$
of the unbiased Edwards-Wilkinson
interface modelled by the symmetric KLS model for any $D^{A0}>0$.
The physics of the biased interface where studied by
Helbing et al \cite{Helb99} by considering asymmetric rates
\bel{4-6}
D^{AY}_{r} = r, \quad D^{0Y}_{r} = q, \quad D^{0Y}_{l} = p,
\quad D^{AY}_{l} = 0.
\ee
with $p=1-q-r$.
For $r=0$, the system satisfies detailed balance with respect to the
measure (\ref{4-4}). Below the wetting transition $q<1/2$ one has the
exact exponents (\ref{4-4}). At the critical point the interface is rough
and one expects a dynamical critical exponent as measured in the related
model of Koduvely and Dhar. Above the critical point the interface would
grow, but cannot detach from the minimal height level due to anchoring.
Hence the measure (\ref{4-4}) is stationary for all $q$. The stationary
interface has a cusp, the anchoring point is random. This is an example
of spontaneous breaking of translational invariance. In the particle
picture the steady state is a shock measure with extremal limiting
densities $\rho_1=\epsilon$, $\rho_2=1-\epsilon$ with a sharp downward shock 
at some random lattice site $k$. In a finite system the small quantity
$\epsilon$ and hence the particle current
is exponentially small in system size $L$. The random
anchoring point $k$ moves with a speed also exponentially small in system 
size. This structure
describes a strongly phase-separated state with an essentially empty
region and an essentially full region. For this to be valid it is not
necessary to require a total average density $\rho=1/2$. It is easy
to understand this steady state directly from the microscopic
dynamics. The right edge of the occupied domain, i.e., the  right-hand shock
is stable because a hopping of the rightmost particle is exceedingly
unlikely since the transition $AA0 \to A0A$ is forbidden and the configuration
$0A0$ where hopping is allowed is exceedingly unlikely for the rightmost
particle (exponentially small in system size). On the other hand, at the
left boundary of the domain the system essentially behaves like an
ordinary ASEP which has a stable upward shock in the direction of motion.

For $r>0$ the interface is not anchored anymore, but the minimal height
condition for any completed layer is still conserved. Along the line
$r=q$ one observes a transition from a rough growing interface (finite
particle current $j>0$ for $q>q_c\approx 0.1515$) with KPZ dynamics
to a smooth interface with spontaneously broken symmetry. The symmetry
breaking can be quantified in terms of the non-conserved order parameter
\bel{4-7}
\tilde{M} = \sum_k (-1)^k n_k.
\ee
This quantity measures the difference of sublattice densities between
the even and odd sublattice respectively. In the language of spin systems
this is the staggered magnetization, playing the role of the order parameter
for antiferromagnetic systems. Below the critical point particles either
accumulate on the even or odd sublattice respectively. Both happens
with equal probability, but a transition between both kinds of configurations
occurs on a time scale that diverges exponentially in system size.
This is the signature of spontaneous symmetry breaking in a finite system.
An explicit calculation of the exponentially large transition time 
is possible in the vicinity of the line $q=0$ \cite{Schuunpub}.
No reference to the height variable is necessary for measuring 
$\tilde{M}$. Numerical investigation of the model shows that it is
in the same universality class as the growth model of Alon et al.
\cite{Helb99}.

In neither of the models discussed above the positive-rates condition is 
satisfied. Only kinetic constraints imposed
by vanishing rates, analogous to
the zero-temperature condition on phase transitions in equilibrium,
may lead to a vanishing current in which case the stability argument
for domain walls does
not apply and phase transitions can occur.
Below we shall present an independent
argument that suggests conditions under which kinetic constraints
lead to phase separation.
Hence so far there is no known simple model with a single 
conservation law that violates the positive rates conjecture.

\subsection{Phase separation in two-species ASEP's}

The exact and numerical analysis of steady states of one-species systems
reviewed above has revealed that phase separation in systems defined on a 
finite ring or on $\bbbz$ may occur if one or more of the following 
conditions are satisfied:
\begin{itemize}
\item (I) there are spatially localized defects reducing the mobility of 
particles
\item (II) single particles of a different species act as mobile blockages
\item (III) the dynamics have kinetic constraints arising from a 
nonequilibrium 
zero-temperature condition.
\end{itemize}
The last condition leads to {\it strong phase separation} in the sense
that one domain is fully occupied whereas the other domain is entirely
empty. The current in the phase separated state vanishes exponentially
in the size of the particle domain, the separated
state exists at any total particle density. 
Conditions (I) and (II) may lead to strong phase separation, but
allow also for a soft phase separation between domains of different
densities. This phenomenon sets in only for densities above some
critical density $\rho_c$. The steady-state current is nonvanishing
and independent of $\rho$ in the phase separated state: Increasing the
density leads to an increase of the size of the high-density domain,
but not to a change of the current. In analogy to Bose-Einstein condensation
we call the high density domain a condensate, the transition at $\rho_c$
is referred to as condensation transition.
Notice that this characterization refers to the thermodynamic limit
$L\to\infty$. In a finite system there is either a current exponentially
small in system size (case A, strong phase separation) or one has
finite-size corrections to the finite bulk current (case B, soft phase 
separation). We remind the reader that soft phase separation may disappear 
above a critical density $\tilde{\rho}_c$ and also for a finite density
of blocking particles \cite{Lee97}. Strong phase separation is accompanied
by spontaneous breaking of translational invariance, except if caused
by condition (I) where translational invariance is explicitly broken.

Strong phase separation has been found also in homogeneous systems on a ring
where neither of the conditions (I) - (III) is satisfied, but where there
is a second species of particles with finite density 
\cite{Lahi97,Arnd98b,Evan98a}. Hence we add a further
sufficient condition for the possibility of phase separation
\begin{itemize}
\item (IV) the system has two or more conservation laws
\end{itemize}
We remark that all the conditions (I) - (IV) in some way or other impose
local constraints on the dynamics of the driven diffusive system. This
appears to be a general requirement for phase separation in generic
driven diffusive systems. The size of the local state space and the
range of interaction appear to be irrelevant if one of the conditions
(I) - (IV) is satisfied.

Using a four-states model which is equivalent to a two-lane
model with two conserved densities Lahiri and
Ramaswamy \cite{Lahi97,Lahi00} address the question of phase separation in 
terms of the stability of crystals moving steadily through a dissipative 
medium, e.g., a sedimenting colloidal crystal. In a certain limit
(large particle radius or small elastic modulus of the suspension) experiments
suggest instability of such a crystal. Numerical analysis of the lattice
model, however, reveals a transition to a stable regime, corresponding
to strong phase separation. In the simpler 2-species ASEP (\ref{3-11}) 
or in a more symmetric model with rates \cite{Evan98a,Evan98b}
\bel{4-8}
D_{A0}=D_{0B}=D_{BA}=1,\quad D_{0A}=D_{B0}=D_{AB}=q
\ee
the mechanism for strong phase separation for $q<1$ is very transparent.
Here strong phase separation refers into separation of three pure
macroscopic domains, each consisting of essentially only one particle
species or empty sites.
For simplicity we assume $N^A=N^B$, but this is not necessary for the
phenomenon to occur. Prepare a phase-separated block 
which we symbolically represent
by $\dots 000AAAAAABBBBBB000 \dots$. One observes the following: (i) The
$0|A$ interface is stable by the criterion (\ref{3-21}) since due to the
absence of $B$-particles one has the dynamics of the usual ASEP (with 
a bias to the right) in the
vicinity of this domain wall. (ii) The
$B|0$ interface is stable for exactly the same reason ($B$ particles have
a bias to the left) (iii) The $A|B$ interface is stable since in the
absence of vacancies $B$-particles act like vacancies w.r.t. the
local dynamics of the $A$-particles and vice versa. (iv) Since each
domain wall is stable (only small fluctuations extended
over a finite range of lattice sites evolve at the phase boundaries)
the assumption used in the argument remains valid
for all times. 

It is clear that this model can be extended to an arbitrary number
of conserved species and does not require equal density for each particle
species. However, for $N^A=N^B=L/3$ the dynamics can be shown to satisfy
detailed balance w.r.t. an equilibrium measure with a non-local interaction
energy of Ising type \cite{Evan98b}
\bel{4-9}
E = \sum_{k=1}^{L-1} \sum_{l=k}^L
\left[(1-n^A_k-n^B_k)(n^B_l-n^A_l) + n^A_k n^B_l \right]
\ee
The corresponding partition is proportional to system size $L$, rather
than $\exp{(fL)}$, since fluctuations occur only in a finitely
extended region around the three domain walls. The ``temperature''
associated with equilibrium measure is given by $kT = -1/(\ln{q})$.
It diverges at $q=1$ which corresponds to the disordered state
of symmetric diffusion of first- and second-class particles.
The analysis of the model for $q>1$ is similar, with the role
of $A$ and $B$-particles interchanged. The behaviour of the model
close to the transition point $q=1$ has been investigated \cite{Clin02}
and some critical exponents have been determined numerically.

The related but distinct model (\ref{3-11}) of Arndt et al. has strong
phase separation for $q<1$, with essentially nonfluctuating $0|A$ and
$B|0$ domain walls, and an $A|B$-interface similar to the one in
model (\ref{4-8}). For $q>1$ the behaviour of the model is more intricate.
Numerical and mean-field analysis \cite{Arnd99} suggests the existence of soft
phase separation up to a critical value $q_c=1+4\lambda\rho/(1+2\rho)$.
There is a condensate of density 1, but consisting of both species of 
particles, but vacancies. The other ``fluid'' phase has density $<1$, with 
particles of both species and vacancies
distributed apparently similarly to the disordered phase for $q>q_c$. 
Notice that inside the condensate particles also flow, but with the
dynamics of the usual ASEP, as the $B$ particles act like vacancies
in the usual ASEP. The number of $A$ and $B$ particles is on average
equal and the condensate essentially behaves like the usual
ASEP with open boundaries in the maximal current phase, except that
the system size $M$ of this ASEP corresponds to the slightly fluctuating 
cluster size. The stationary current for both particle species
is non-zero and approximately
given by the value  $J=(q-1)/4$ expected from the ASEP in the
maximal current phase. The position of the condensate fluctuates
on the lattice. For an unequal average density of $A$ and $B$-particles,
the scenario as described here remains essentially unchanged,
except that the condensate has a finite drift velocity \cite{Arnd02}.

The exact stationary distribution of the model (\ref{3-11}) can be calculated
using the matrix product ansatz. For the grandcanonical ensemble
with equal densities of both particle species it has been shown \cite{Raje00} 
that
the apparent condensation transition is a crossover effect. For sufficiently
large lattice one would observe a distribution of clusters, but not a
single macroscopic condensate. Using the parameter 
$a=(1-q^{-1}-\lambda)/\lambda$
and the fugacity $\xi$ controlling the density the exact current $J$ and
the density $\rho$ are given by
\bea
\label{4-10}
J(\xi) & = & \frac{2a^2\xi}{1+a^2+2a(1+a^2)\xi - (1-a^2)\sqrt{1+4a\xi}}\\
\label{4-11}
\rho(\xi) & = & \frac{a(1+a)\xi\left[(1+a)\sqrt{1+4a\xi}-(1-a)\right]}
       {\sqrt{1+4a\xi}\left[1+a^2+2a(1+a^2)\xi - (1-a^2)\sqrt{1+4a\xi}\right]}.
\eea
This expression has a very remarkable property: Inside the 
apparently condensed phase at $q=-a=10/9$ the derivative 
$J'(\rho)$ which enters the Jacobian (\ref{3-22}) and hence determines
the collective velocities of the two-species system has a change of
order 1 arising from a change of order $10^{-24}$. One would need
a lattice of the order of $10^{70}$ sites to actually observe the
breakdown of the condensation and see the full distribution of clusters
of various lengths.

Clearly, a crossover scale of this magnitude is of no relevance for the
occurrence of soft phase separation in a real system. Any finite sample
would exhibit a phase separated state. However, the huge number $10^{70}$ 
characterizing the typical scale of the cluster
size distribution is specific for the model (\ref{3-11}). 
As shown in \cite{Kafr02a}
the crossover scale depends sensitively on nonuniversal parameters
which are tunable in some models \cite{Kafr02b}. Hence other models
may have parameter ranges with a mean cluster sizes of the order, 
say, $10^{5}$. Such a crossover
scale would render numerical results from computer simulations of 
the steady state for a real
quasi one-dimensional system ambiguous:
A computer simulation
of a realistic model for a real system with less than
$10^5$ particles
could predict soft phase separation, whereas actual experiments done
on a macroscopic sample with more than $10^5$ particles
could yield the contradictory result that
there is only a disordered phase. Yet, one could not conclude from this
observation that the
model is inappropriate to describe the real system since in a smaller
experimental sample of the same system computer simulations and experimental
observations may agree.

We remark that a technical assumption in the exact calculation of
Rajewsky et al \cite{Raje00} has been proved in \cite{Sasa01}. The
validity of the result also in the canonical ensemble has been
challenged \cite{Arnd02}, even though exact analysis of the fluctuations
in the particle density strongly suggest that there is no true condensation
in the canonical ensemble \cite{Raje00}. In another model
introduced by Korniss et al a two lane extension
of a 3-species driven system was studied~\cite{Korn99,Mett02}. It has
been suggested that while for this model the one lane system does
not exhibit phase separation~\cite{Godr}, this phenomenon does
exist in the two lane model. The studies rely on numerical
simulations of systems of length up to $10^4$. However, no
theoretical insight is available as to why phase separation in this
two-lane model should persist in the thermodynamic limit.

To conclude these general considerations we note that in many real quasi
one-dimensional systems the particle number
is in the range $10 \dots 10^4$. Simplified models of such systems
are accessible to numerical simulations of the steady state.
This raises the further question to which extent phase separation in 
finite systems is obscured by a too large intrinsic width of the
domain wall separating the condensed domain from the ``fluid''
low-density
domain. While quantitative predictions for the relaxation
modes resulting from the
coarse-grained domain wall theory in single-species systems with open
boundaries have been verified numerically for the ASEP on small
lattices of only
${\cal O}(10)$ sites \cite{Nagy02} there is no systematic finite-size study
of soft phase separation.

\subsection{Criterion for phase separation}

In view of the exact analysis of the model (\ref{3-11})
it is clear that numerical evidence for
soft phase separation may be rather subtle and indeed be misleading.
It would thus be of great importance to find other criteria, which
could distinguish between models supporting phase separation from
those which do not. Phase separation is usually
accompanied by a coarsening process in which small domains of,
say, the high density phase coalesce, eventually leading to
macroscopic phase separation. This process takes place as domains
exchange particles through their currents. When smaller domains
exchange particles with the environment with faster rates than
larger domains, a coarsening process is expected, which may lead
to phase separation.

An approach that quantifies this mechanism and yields a criterion
for phase separation in terms of the current leaving the domains
is proposed by Kafri et
al \cite{Kafr02}. The current-criterion
is readily applicable even in cases which cannot be decided by
direct numerical simulations. In order to explicitly state the criterion 
one distinguishes systems with a vanishing
current of a finite domain of size $n$ 
\bel{4-14}
J_n \to 0 \quad \mbox{(case A)}
\ee
from systems finite-size corrections to a finite asymptotic
domain current $J_\infty$ of the form 
\bel{4-15}
J_n = J_\infty (1+b/n^\sigma) \quad \mbox{(case B)}.
\ee 
to leading order in $1/n$. For simplicity we assume here domains
with vanishing drift velocity in which case the current inside the
domains equals the outgoing current. More generally one has to
distinguish the two currents leaving the cluster at the right and
left boundary respectively.

For $b>0$ the current
of long domains is smaller than that of short ones, which leads to
a tendency of the longer domains to grow at the expense of smaller
ones. The current criterion asserts that phase separation exists only
in the following cases \cite{Kafr02}: 
\bea
\label{critA}
J_n & \to & 0 \quad \mbox{\rm for $n\to\infty$ (case A)} \\
\label{critB}
J_n & \to & J_\infty>0 \quad \mbox{\rm (case B)}
\eea
for either $\sigma<1$ and $b>0$ or $\sigma=1$ and $b>2$.
In case A one has strong phase separation for any density, whereas in case B 
one has soft phase separation at any density for $\sigma<1$ and 
above a critical density
\bel{4-16}
\rho_c = \frac{1}{b-2}
\ee
for $\sigma=1$. The fluid regime has particles with density
$\rho_c$. Hence in a finite system the macroscopic size of the condensate in 
the phase-separated regime is determined by the 
system parameter $b$. For an asymptotic decay faster than $2/n$
there is no condensed phase, the system is disordered for all densities.

Models with two conservation laws 
for which $J_n$ decays exponentially to zero with $n$ 
(case A) have been
reviewed above and indeed were shown to exhibit strong phase
separation at any density. In the model (\ref{4-6}) with a single 
conservation law
the current of particles out of the left domain
wall, i.e., the current opposite to the bias of the
individual particles, decays exponentially with domain size for $r=0$, 
as demonstrated for the usual ASEP with appropriately chosen open boundary 
conditions
\cite{Blyt00}. The domain size dependence of the
current flowing away from the right edge
of a cluster (in direction of the bias)
is to leading order not a self-organized
quantity, it is determined strongly
by the interaction of the particle at the edge of the cluster with the
surrounding particles. The interaction range is one lattice site
and hence the current $J_n$ out of the right edge of a cluster of density 1 
becomes to leading order independent
of the cluster size for $n>2$. It vanishes due to
the kinetic constraint $r=0$ (zero-temperature condition). According to the 
criterion one expects strong phase separation, in agreement with
the result reviewed above. 

For  $J_\infty \neq 0$ (case B) we note that
in a system with two conservation laws the current inside a cluster
organizes itself to a value determined by the dynamics of the
reduced system with only one conservation law resulting from the absence of 
vacancies. This reduced system has open boundaries 
with in- and outflow of particles such that the system is in the
generic maximal current phase of the reduced system.
It is assumed that the current flowing through a block is
given by its steady-state value and is independent of its
neighboring blocks. This may be justified by the fact that the
coarsening time of large domains is very long, and the domains
have a chance to equilibrate long before they coarsen.

In case B one expects generically $\sigma=1$ for the following
reason: (a) In a {\it periodic system} the leading finite-size 
corrections to
the current $J_\infty$ in a canonical ensemble is given by
$J_n - J_\infty = - J_\infty''\kappa/(2n)$ \cite{Krug90b,Kafr02b}. 
Here $J_\infty''$ is the
curvature of the current-density relation and $\kappa=(\exval{N^2}-
\exval{N}^2)/L$ is the nonequilibrium analog of the thermodynamic
compressibility which is assumed to be finite, i.e., one assumes
sufficiently rapidly decaying correlations as was implied above in the
derivation of the collective velocity which also requires finite
compressibility. (b) There is a universal ratio $c^\ast$
of the finite-size corrections to the current in the maximal
current phase of a driven diffusive system (which describe the
dynamics inside the growing domains) and the finite-size corrections
of the canonical ensemble of a periodic system \cite{Krug94}.
This yields leading finite-size corrections of the form (\ref{4-14})
with a parameter $b$ entirely determined by the universal constant
$c^\ast$ and the macroscopic quantities $J_\infty''$ and $\kappa$.
The value of $c^\ast=3/2$ has been obtained from the exact solution of the
ASEP with open boundaries \cite{Schu93b,Derr93b}.

We stress that by definition $b$ is a quantity that itself does not depend on
system size. For systems with unknown
stationary distribution the reduced dynamics inside a cluster
allows for a simple numerical measurement
of $b$ by studying the finite-size corrections of the stationary current
in the reduced open system of length $n$. One neither needs huge lattices nor
is one faced with the problem of slow relaxation of the phase separation
process in the full system. Applying the criterion to the model 
(\ref{3-13}) yields the exact value $b=3/2$ and hence
one expects no condensation, in agreement with the exact result. 
For the two-lane model of Korniss et al 
\cite{Korn99} one
obtains numerically
$b\approx 0.8$ \cite{Kafr02} and therefore one expects no condensation in
contrast
to the results of the Monte-Carlo simulation of the full model
with $10^4$ lattice sites. A three-states model with KLS-dynamics
(\ref{2-9a})
inside the clusters has been shown to have $b>2$ \cite{Kafr02b}
which suggests the existence of soft phase separation in driven
diffusive systems with two conservation laws.

The criterion presented above emerges from a careful analysis of the
zero-range process (ZRP) which could be viewed as a generic model
for domain dynamics in one-dimension \cite{Kafr02}.

Depending on the rates $w_n$
the model may or may not exhibit condensation in the thermodynamic
limit, whereby the occupation number of one of the boxes becomes
macroscopically large. Clearly the rate $w_n$ must be a decreasing
function of $n$ in order for larger blocks to be favored and to
support condensation. It is known~\cite{Evan00,Oloa98} that
condensation occurs at any density when $w_n \to 0$ with $n \to
\infty$, or when it decreases to a non-vanishing asymptotic value
as $b/n^\sigma$ with $\sigma<1$; no phase separation takes place
for $\sigma>1$; for $\sigma=1$ phase separation takes place at
high densities only for $b>2$.
This model may be used to gain physical insight into the dynamics
of driven one-dimensional systems. Occupied boxes represent
domains of the high density phase. The currents leaving domains
are represented by the rates of the ZRP. This is done by
identifying the rate $w_n$ associated with a box containing $n$
balls with the currents $J_n$ leaving a domain of $n$ particles. A
bias in the currents to a certain direction may be incorporated
through a bias in the ZRP dynamics. The existence of a box with a
macroscopic occupation in the ZRP corresponds to phase separation
in the driven model. The distribution of occupation numbers
obtained from the ZRP was shown to agree with the domain size
distribution of the model (\ref{3-13}) \cite{Kafr02}.

It is remarkable that extending the asymptotic expansion (\ref{4-15})
by a (nonuniversal) next-leading term $c/n^2$ leads to an extremely
sensitive dependence of the mean domain size $\xi$ on $c$ \cite{Kafr02a}.
The quantity $\xi$ exhibits a sharp
increase of a few orders of magnitude over a narrow range of
values of $c$. This reflects itself in large (but finite) blocks
and an apparent phase separation in direct simulations.

\subsection{Coarsening}

The previous discussion has focussed on the stationary properties
of phase separation. The dynamics below the critical density are
expected to be described on the hydrodynamic scale by the mechanisms
reviewed in Sec. 3. Above the critical density a natural set-up is to 
start with particles 
uniformly distributed at the supercritical density $\rho >\rho_c$. 
In the beginning the excess particles condense at a few random sites. 
Thus there are several clusters which are essentially immobile. 
On the remaining sites, the distribution relaxes to its critical
stationary distribution with $\rho =\rho_c$. With increasing time the 
larger clusters 
will gain particles at the expense of the smaller ones, causing the 
clusters to merge. Eventually only a single cluster 
containing all excess particles survives, which is 
typical for the stationary distribution (Fig.~6).

In case A (strong phase separation) the evolution of this state will
proceed by slow diffusion against the bias, in which, for example
an $A$-particle traverses the adjacent domain of $B$-particles.
Using standard mean first passage time calculations the time $t$ necessary
for penetrating the complete $B$-domain of size $n$ can be shown
to be of order $q^n$ where $q>1$ is the hopping asymmetry, i.e., the
ratio of hopping rates inside the $B$-domain. After that the $A$-particle
travels with finite average velocity to the edge of the next
$A$ domain. This happens after a time of order $L$ which can be neglected
compared to the penetration time. Therefore the coarsening time
is of the order $q^n$ and inverting this relation yields 
a logarithmic growth law
for the mean domain size in the strong phase separation \cite{Evan98b}
\bel{4-12}
\bar{n} (t)\sim \ln{t}/\ln{q}.
\ee
This growth law is valid also in higher dimensions \cite{Kafr00}.

The zero-range picture may be used in order to study the coarsening
dynamics of domains in the case of soft phase separation.
We set $J_\infty=1$ which only fixes
an uninteresting microscopic time unit for coarsening.
 In \cite{Gros03b} 
the coarsening dynamics are described by studying with random walk
arguments the loss and
gain of particles in neighbouring clusters, mediated by stationary transport
in the fluid phase in between. The time that particles lost from a
cluster spend in between clusters is of the same order as the time 
required to dissolve a cluster completely. Hence this time scale
is the appropriate coarsening time scale. In the totally asymmetric
case excess particles leave a cluster site with $n$
particles at a rate $b/n$ and therefore
the typical required to dissolve such a cluster is given by
$t_a (n)\sim {n^2/b}$. Inverting this scaling relation
yields the growth law 
\bel{4-17}
\bar{n} (t)\sim (bt)^{1/2}
\ee
for the mean cluster size $\bar{n}$. From (\ref{4-17}) one reads off
the dynamical exponent $z=2$ in the asymmetric condensed phase.

Because of the recurrence of a 1-d random walk 
in the case of symmetric hopping, a particle that got lost from a
cluster is very likely to return to the 
cluster it left, in contrast to the asymmetric case. It 
will be trapped by the next cluster only with a probability
inversely proportional to  the diffusion distance, 
i.e., of the order $(\rho -\rho_c )/m$. So the typical time of a particle 
to leave a cluster is of the order $m^2 /(b(\rho -\rho_c ))$. Hence the
coarsening time scale is given by 
$t_s (n) \sim t_a {(\rho -\rho_c )/m}\sim {m^3 /[b(\rho -\rho_c )]}$.
This results in the growth law
\bel{4-18}
\bar{n} (t)\sim [b(\rho -\rho_c )t]^{1/3}
\ee
and dynamical exponent $z=3$ in the symmetric case. A similar growth law
which can be rationalized using analogous arguments is known for 
equilibrium Kawasaki dynamics (\ref{2-9b}) at very low temperatures
\cite{Corn91,Maju94}. For biased coarsening dynamics with conserved
order parameter one has $z=2$ \cite{Corn96,Spir99},
in agreement with  (\ref{4-17}). The coarsening
stops when only one macroscopic cluster is left.
The typical time of a macroscopic fluctuation of the cluster size 
diverges exponentially with the system size $L$ \cite{Gros03b}.

The growth laws (\ref{4-17}), (\ref{4-18}) were confirmed by Monte-Carlo
simulations of the ZRP \cite{Gros03b}. The critical exponents were
obtained independently by numerically studying the second moment
of the local density \cite{Godr03}. Also a universal scaling function
for the cluster size distribution was obtained. At the critical density
itself Monte Carlo simulations suggest dynamical exponents
$z_c \approx 3$ for the asymmetric case and $z_c \approx 5$ for the
symmetric case. A theoretical derivation of these exponents is lacking.

\section{Conclusions and open questions}
\setcounter{equation}{0}

During the last decade the study of one-dimensional driven diffusive systems
has contributed significantly to the understanding of critical phenomena
far from equilibrium, where ``understanding'' not only refers to the
characterization of nonequilibrium universality classes in terms
of critical exponents, but also to the
identification of some of the coarse-grained dynamical mechanisms that 
generate
these critical phenomena. Exact results for simple model
systems such as the ASEP have played a crucial role in advancing and shaping
this understanding. They have
provided deep and detailed insights
in robust generic phenomena which could then be generalized to more 
complicated 
systems.

This includes the derivation of exact hydrodynamic
equations and asymptotic coarsening laws
from the microscopic laws of interaction, both rigorously and
using more intuitive physical arguments based on the application of
conservation laws and using random walk arguments. Thus complicated
collective
phenomena such as shocks can be viewed as effective single-particle
excitations with simple properties. Universal fluctuations determining
the dynamics on intermediate scales between the microscopic lattice scale
and the macroscopic Euler scale have been probed using test particles:
tracer particles, second class particles, impurity particles. Thus some
similarities to the still unresolved problem of localized blockages
could be established. The occurrence of stationary critical phenomena,
viz. spontaneous symmetry breaking, long-range order and phase separation
of two types, soft and strong
could be linked to microscopic properties of the dynamics, listed
in conditions I - IV of Sec. 4. Stability criteria (\ref{3-21}), 
(\ref{3-23}) for microscopically
sharp domain
walls and coarsening dynamics of domains (\ref{critA}), (\ref{critB})
as well as the phase diagram of open one-species systems (\ref{min})
can be phrased directly in terms
of the macroscopic current. We conclude that the current-density
relation (which is determined by microscopic parameters)
is a central quantity determining the large scale physics of
driven diffusive systems. This is fortunate in the sense that the
current is usually relatively easy to measure or calculate. It is crucial to
know the current exactly, mean-field approximations are not likely
to produce even qualitatively (let alone quantitatively) correct 
features of the system, unless by chance certain analytic properties
of the mean field current happen to coincide with those of the
exact current. However, this is not guaranteed even if bulk correlations
in the stationary state are short ranged. 

A major open question concerns the precise relationship of the various
current criteria and the microscopic criteria for phase separation.
Addressing this issue leads to series of interrelated problems, each
of which is interesting in its own right.

It is not clear how to deduce the stability of a domain wall in the
stationary condensed state or during the coarsening process 
(where the current is also stationary and independent of the density)
from the stability criteria (\ref{3-21}), (\ref{3-23}). This requires
a more careful analysis of the behaviour of fluctuations in systems
with two conservation laws and is intimately linked to
a hydrodynamic description of the dynamics above the critical density,
which is an open problem even within  the framework of an effective
ZRP description. The hydrodynamic treatment of the blockage problem
in the usual ASEP \cite{Baha98} may provide some insight. 

Moreover it would be highly desirable to have a unified picture which allows
for an application of the current criteria (\ref{critA}), (\ref{critB})
for phase separation to
single-species systems satisfying one of the conditions (I) - (III)
and thus to predict from the dependence of the domain current
on the microscopic parameters at which value of these parameters
phase separation sets in. This requires a proper definition of the
current out of a domain in system with one conservation law which
might then answer also the 
question whether soft phase separation is possible if neither of the conditions 
(I) - (III) are satisfied. In a single-species system
this condensation phenomenon is reminiscent of spontaneous traffic jams
in automobile traffic flow.  Indeed, traffic models with nonconserved 
internal degrees of freedom are known to exhibit soft phase separation
\cite{Taka93,Scha97,Benj96b,Barl98}, but the minimal requirement
for the existence of the phenomenon is not entirely clear. 
The role of non-conserved internal degrees of freedom in critical 
phenomena needs further clarification also in
the theory of boundary-induced phase transitions which requires
some extension \cite{Nama02}. 

In this context it would be interesting to try to predict phase separation
from the properties of the current, using  (\ref{3-21}), (\ref{3-23}).
This has not been attempted yet even in systems where both the current
and the existence of phase separation are known and
might shed light on the possibility
of phase separation between domains of nonextremal densities $\rho_i \neq 0,1$
in translation
invariant systems. 
The stability criterion (\ref{3-21})
rules this out for generic current-density relations in systems with one
conservation law, but it is not tantamount to a no-go theorem as a
current which is constant in some density range may allow for such phase
separation.

A further promising and closely related direction of research concerns the 
hydrodynamics of systems with more than one conservation law.
The lack of a full hydrodynamic description of particle systems with two
conservation law still constitutes a major gap in understanding
both dynamic and critical stationary critical phenomena. The selection
of the physical solution using a regularization by adding a phenomenological
viscosity term is not fully understood, as the nature of such a
term might not be as arbitrary
as for systems with one conservation law. This may be of importance
not only for the bulk critical phenomena reviewed here, but also for
boundary-induced spontaneous symmetry breaking \cite{Evan95} and
steady state selection \cite{Popk01} in open systems. There is no
theory of boundary-induced phase transitions with an extremal
principle analogous to (\ref{min}) that could explain
quantitative features of the phase diagram in terms of effective
boundary densities as is possible for single-species models.
Hence it is difficult to make predictions of the stationary
behaviour in open systems. For systems with more than two
conservation laws the problem is likely to be even more intricate,
but possibly also even more fascinating.

Universal properties of fluctuations which manifest themselves
on scales below the Euler scale are not readily accessible with
the analytical methods reviewed in this article. However, with the tools
of Bethe ansatz and random matrix theory they have 
become amenable to exact treatment in the framework of the ASEP
\cite{Prae02,Joha00}. It is natural to try to
apply these techniques to the general 2-species ASEP, or at least to the 
integrable cases. Universal quantities --
critical exponents and scaling functions, but also universal
dynamical mechanisms such as evolution of shocks and coarsening --
derived from  studying driven diffusive systems are ultimately expected
to be useful also in the investigation of real systems such as
those listed above. Therefore one needs to understand the
role of the lattice in the phenomena discussed above. Passing to driven
diffusive system defined in spatial continuum by taking appropriate
limits may give an answer.

\newpage
\subsection*{Figure Captions:}

\noindent Fig. 1: Asymmetric simple exclusion process and related zero-range process. In the ASEP
particles on a lattice hop with rates $D_{r,l}$ to the right and left respectively,
provided the target site is empty. In the case of open boundaries they are created
or annihilated with rates $\alpha,\beta,\gamma,\delta$ at the boundary sites $1,L$
as indicated in the figure. Reflecting boundaries correspond to $\alpha=\beta=\gamma=\delta=0$, 
in the case of periodic boundaries 
particles may hop between sites $L,1$. In the associated zero-range process
consecutive particles correspond to sites,
the length of empty intervals between them to occupation numbers. Hopping of particle
$k$ from a given in the ASEP corresponds to hopping between sites $k-1,k$ in the ZRP.\\[4mm]

\noindent Fig. 2: Mapping of the ASEP to a restricted solid-on-solid (RSOS) growth model. 
An empty (occupied) site corresponds to a slope +1 (-1) in the associated interface.
Hopping to the right (left) thus is equivalent to a deposition (evaporation) of an atom.
The broken line at the bottom of the interface marks a completed layer. In the absence
of chipping the interface cannot shrink below a completed layer which yields a minimal height
model. For an anchored interface the minimal height of the interface equals the height of
the top-most completed layer. In two-species models the definitions are analogous, but
with allowed local interface slopes $0,\pm1$.\\[4mm]

\noindent Fig. 3: Main features  density profiles in a periodic system in the presence of a localized 
defect or a mobile impurity at site $L$. Below the critical density there is an exponentially decaying
profile behind the disturbance, above the critical density one has a macroscopic ``traffic jam''
extending over a finite fraction of the total length $L$. Depending on the specific system, the
density profile may have some extra structure (not shown here) around the disturbance. The position
of the shock at the end of the high-density domain fluctuates.\\[4mm]

\noindent Fig. 4: Time-evolution of the ASEP on the Euler scale, starting from time $t=0$ (a)
with an unstable shock and a region of positive slope. Any region of positive slope 
evolves into a shock after some time (b) because of the flow of localized perturbations 
(microscopic picture) or characteristics respectively (macroscopic description). The evolution
of the unstable shock is not uniquely defined by the viscosity-free hydrodynamics. Both
evolutions (b) and (c) are solutions of the inviscid Burgers equation for initial state (a).
The physical solution selected by the spatially homogeneous ASEP is shown in (b). The time evolution
shown in (c) corresponds to the ASEP with a defect which is governed by the {\it same} homogeneous
Burgers equation \protect\cite{Baha98}.\\[4mm]

\noindent Fig. 5: Schematic phase diagram with second order phase transition line
(broken curve) between
the disordered phase (A) (growing KPZ-interface) and ordered phase (smooth interface)
(B) with spontaneously broken $Z_2$-symmetry and non-vanishing order
parameter $\tilde{M}$ (\protect\ref{4-7}). At $r=0$, there is a
transition at $q=1/2$ with strong phase separation and
spontaneous breaking of translational invariance (bold line C).\\[4mm]

\noindent Fig. 6: Coarsening of domains during phase separation. The regions between domains have
relaxed to their stationary critical density. Each domain of size $n_r$
has an outflowing current $j_r$ determined by the dynamics inside the domain. In order to have
domain growth the current should decrease with domain size, as indicated in the figure for domains
1,2,3. Domain 2 is expected to be eaten up by domains 1,3 after some time.

\newpage

\subsection*{Figures:}

\begin{figure}[h]
\epsfxsize=0.7\textwidth
\centerline{\epsffile{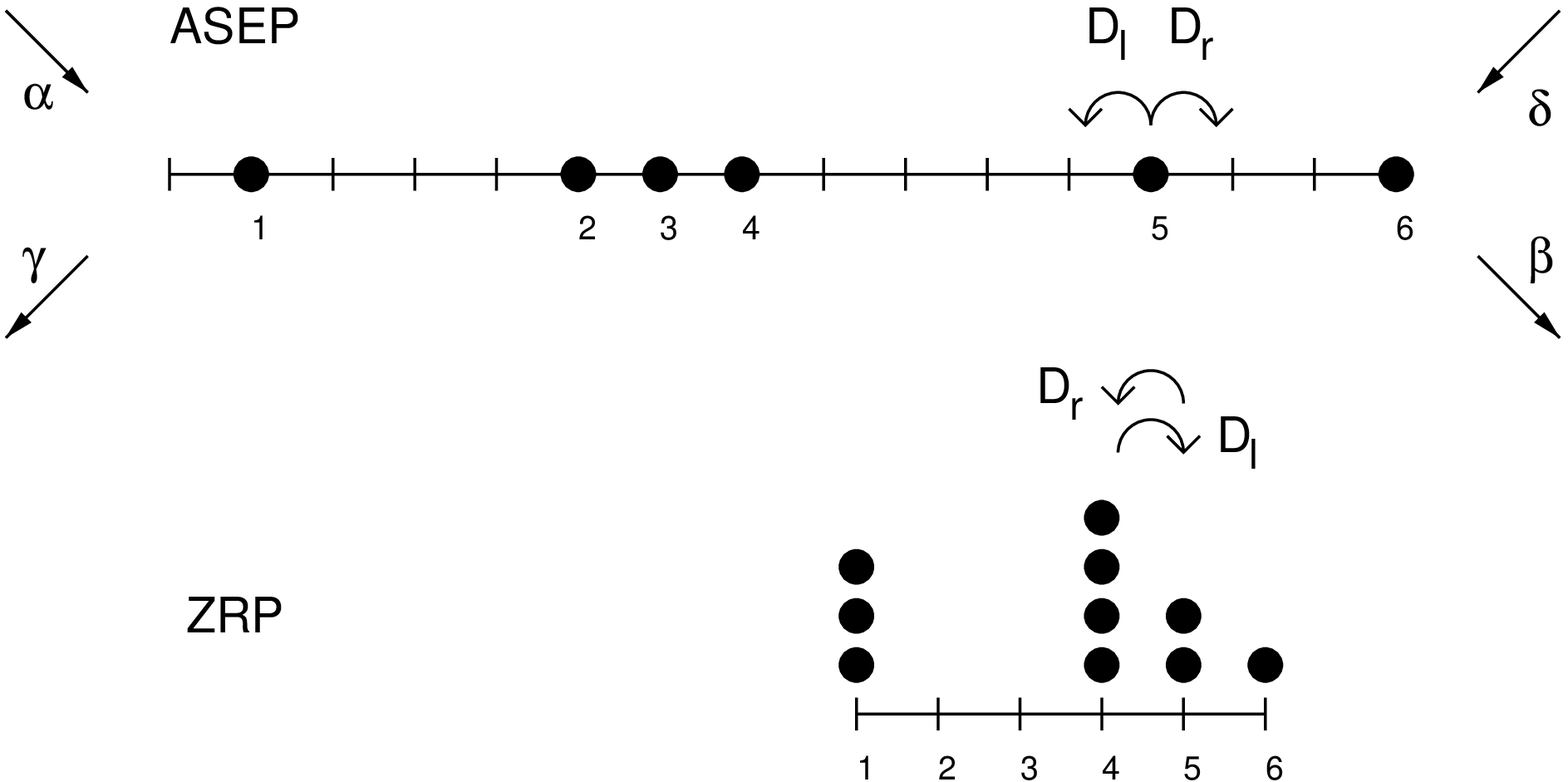}}
\caption{}
\label{Figure1}
\end{figure}
\newpage

\begin{figure}[h]
\epsfxsize=0.7\textwidth
\centerline{\epsffile{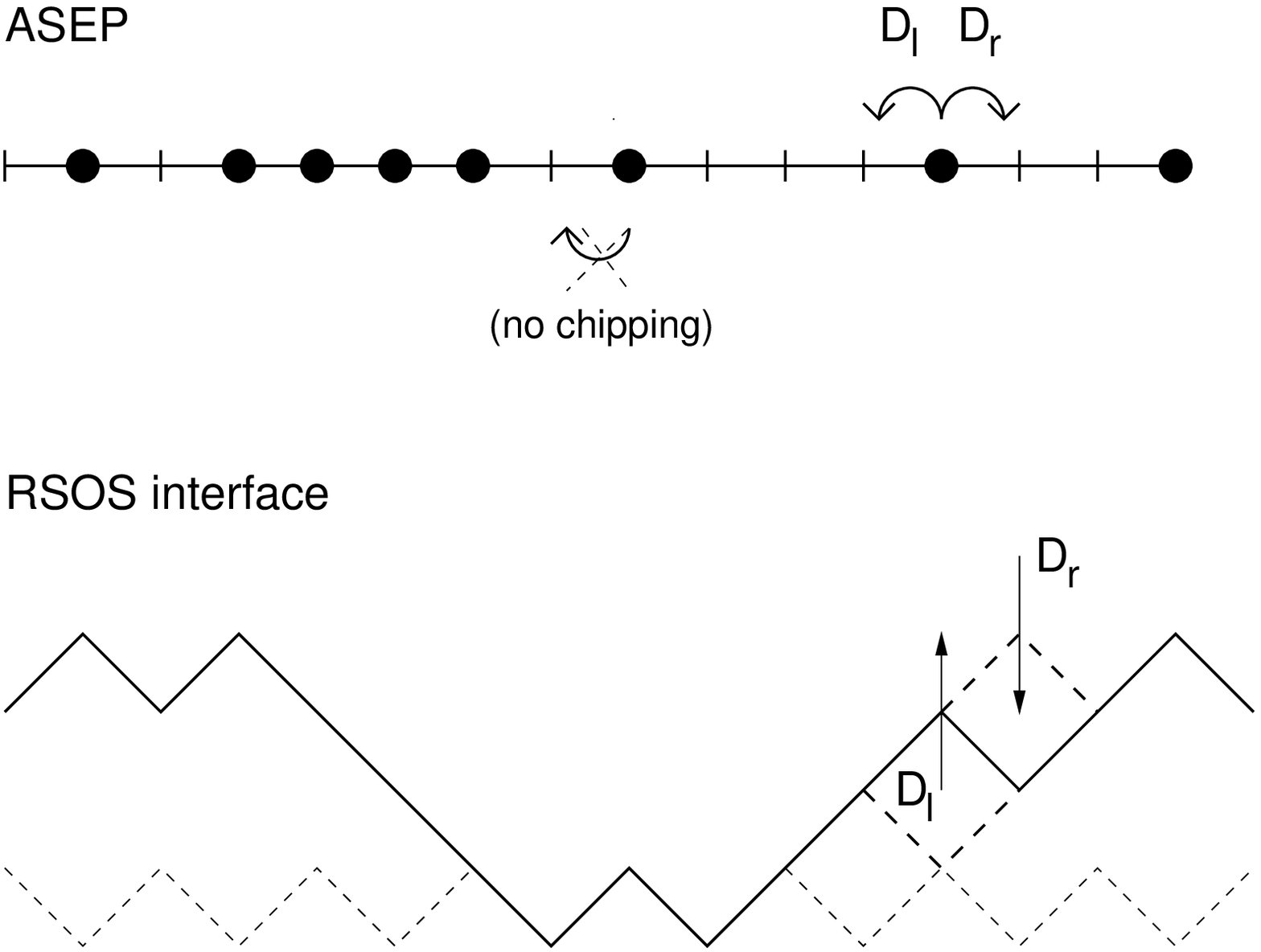}}
\caption{}
\label{Figure2}
\end{figure}
\newpage

\begin{figure}[h]
\epsfxsize=0.7\textwidth
\centerline{\epsffile{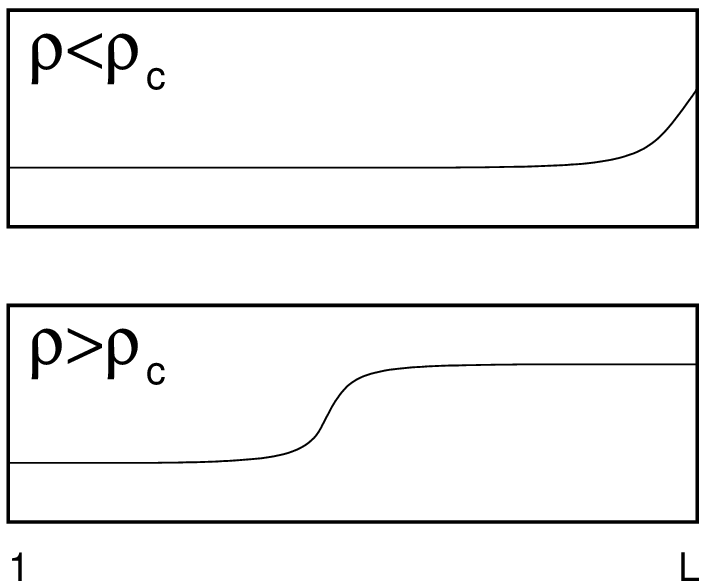}}
\caption{}
\label{Figure3}
\end{figure}
\newpage

\begin{figure}[h]
\epsfxsize=0.7\textwidth
\centerline{\epsffile{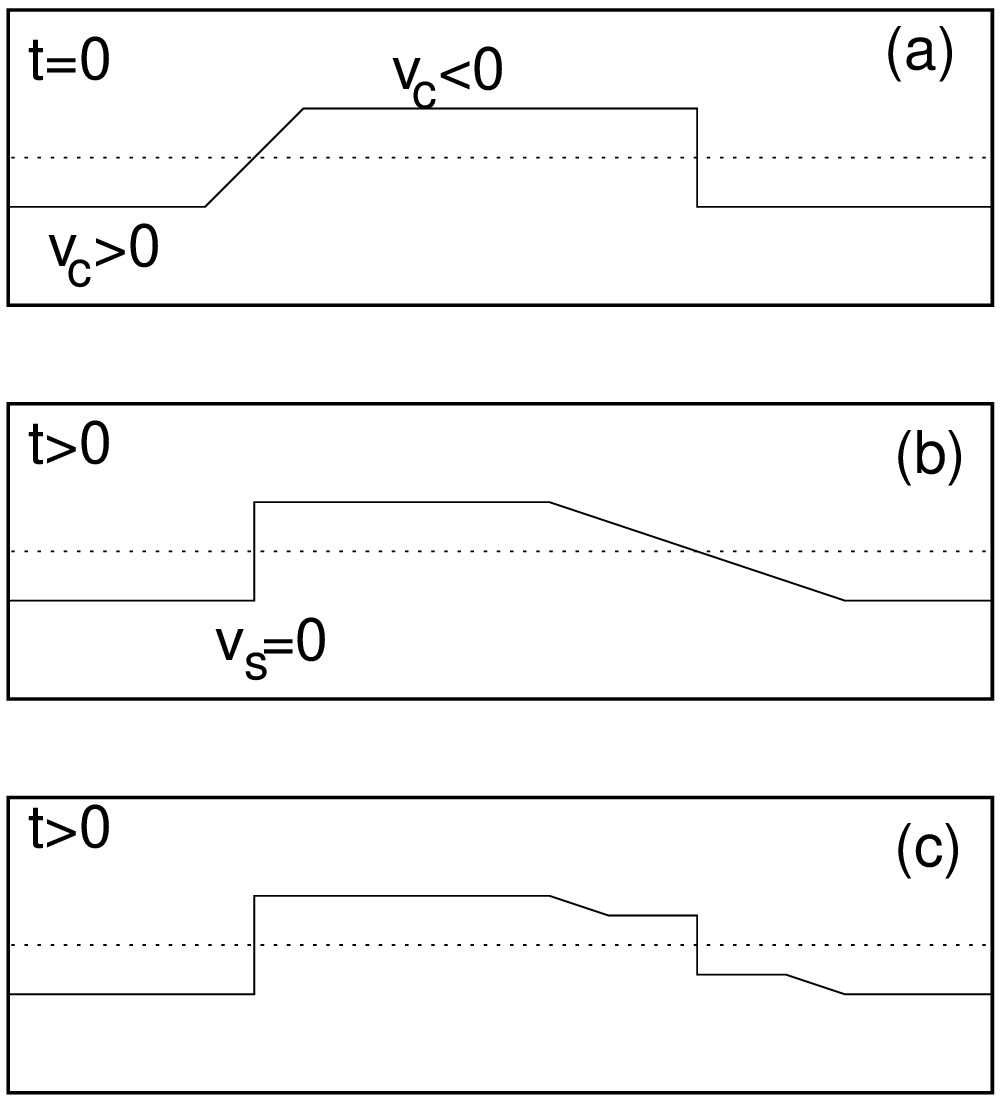}}
\caption{}
\label{Figure4}
\end{figure}
\newpage

\begin{figure}[h]
\epsfxsize=0.7\textwidth
\centerline{\epsffile{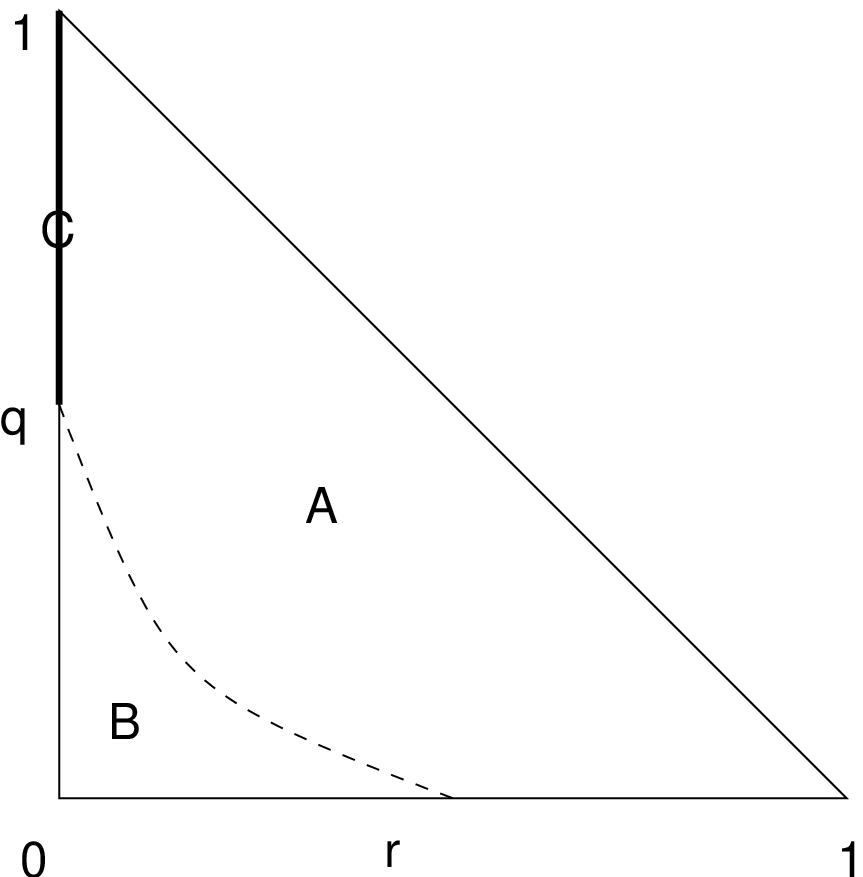}}
\caption{}
\label{Figure5}
\end{figure}
\newpage

\begin{figure}[h]
\epsfxsize=0.7\textwidth
\centerline{\epsffile{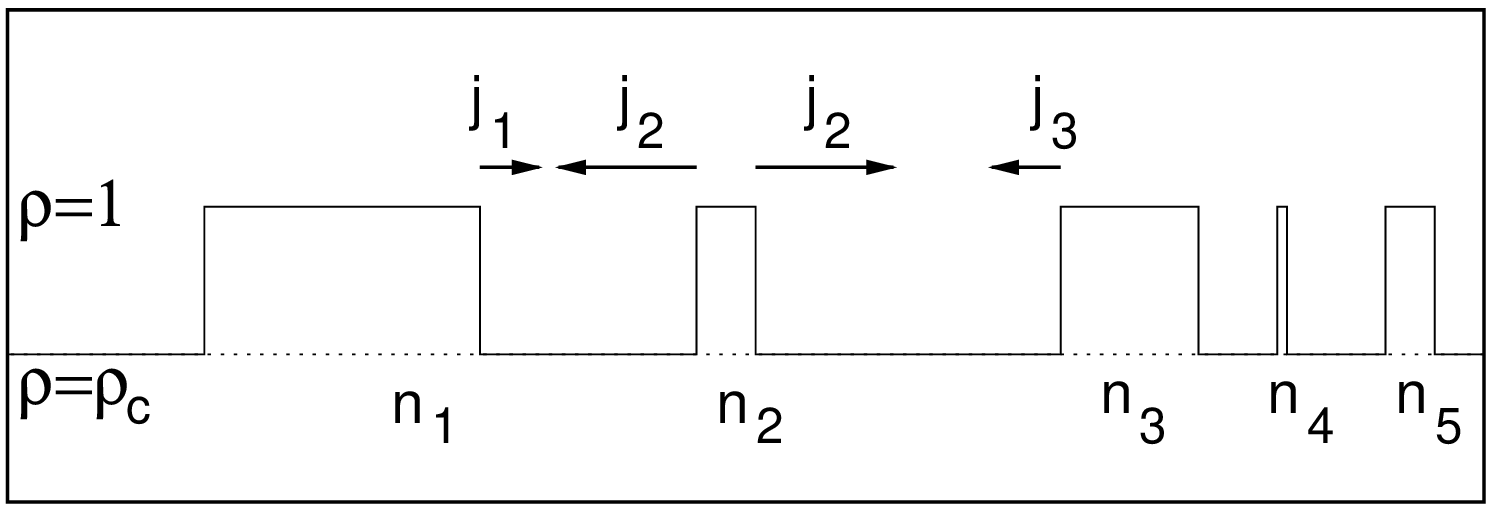}}
\caption{}
\label{Figure6}
\end{figure}


\newpage

\begin{thebibliography}{99}


\bibitem{Kipn99} Kipnis C and Landim C 1999
\newblock {Scaling limits of interacting particle systems}, in:
  {\em Grundlehren der mathematischen Wissenschaften} Vol. 320,
\newblock (Berlin: Springer)

\bibitem{Ferr94b} Ferrari P A 1994 
\newblock Shocks in one-dimensional processes with a drift in: 
{\it Probability and Phase Transition} Ed. G. Grimmett
\newblock (Dordrecht: Kluwer)

\bibitem{Ligg99}  Liggett T M 1999 
\newblock {\it Stochastic Models of Interacting
Systems: Contact, Voter and Exclusion Processes} 
\newblock (Berlin: Springer)

\bibitem{Schu00}  Sch\"{u}tz G M 2001 
\newblock Exactly solvable models for many-body systems far from equilibrium, 
in: {\it Phase Transitions and Critical Phenomena.} Vol. 19,
eds C Domb and J Lebowitz
\newblock (London: Academic Press)

\bibitem{Prae02} Pr\"ahofer M and Spohn H 2002
\newblock Exact scaling functions for one-dimensional stationary KPZ growth
\newblock {\it cond-mat/0212519}

\bibitem{Krug91a} Krug J 1991 
\newblock Boundary-induced phase transitions in driven diffusive systems 
\newblock {\it Phys. Rev. Lett.} {\bf 67}  1882

\bibitem{Popk99} Popkov V and Sch\"utz G M 1999
\newblock {Steady-state selection in driven diffusive systems with open 
boundaries}
\newblock {\it Europhys. Lett.} {\bf 48} 257

\bibitem{Evan00} Evans M R 2000
\newblock Phase transitions in one-dimensional nonequilibrium systems
\newblock {\it Brazilian Journal of Physics}, {\bf 30} 42

\bibitem{Muka00} Mukamel D 2000
\newblock Phase transitions in nonequilibrium systems
\newblock in: {\it Soft and Fragile Matter: Nonequilibrium dynamics, 
metastability and flow} eds M E Cates and M R Evans
\newblock (Bristol: Institute of Physics Publishing)

\bibitem{Jano97} Janowsky S A and Lebowitz J L 1997 
Microscopic models of macroscopic shocks
\newblock in: {\it Nonequilibrium Statistical Mechanics in One Dimension}
ed V Privman
\newblock (Cambridge: Cambridge University Press)

\bibitem{Evan03} Evans M R 
\newblock review in preparation

\bibitem{Schm95}  Schmittmann B and Zia R K P 1995
\newblock Statistical Mechanics of Driven Diffusive Systems, 
in: {\it Phase Transitions and Critical Phenomena.} Vol. 17,
eds C Domb and J Lebowitz
\newblock (London: Academic Press)

\bibitem{Priv97} Privman V (ed.) 1997
\newblock {\it Nonequilibrium Statistical Mechanics in One Dimension} 
\newblock (Cambridge: Cambridge University Press)

\bibitem{Marr99} Marro J and Dickman R 1999
\newblock Nonequilibrium phase transitions in lattice models
\newblock (Cambridge: Cambridge University Press)

\bibitem{Hinr00} Hinrichsen H 2000
\newblock Nonequilibrium critical phenomena and phase transitions into
absorbing states
\newblock {\it Adv. Phys.} {\bf 49} 815

\bibitem{Ziel02}  Zielen F and Schadschneider A 2002
\newblock Exact mean-field solutions of the asymmetric random average process
\newblock {\it J. Stat. Phys.} {\bf 106} 173

\bibitem{Ligg85}  Liggett T M 1985 
\newblock {\it Interacting particle systems} 
\newblock (Berlin: Springer)

\bibitem{Sand94a}  Sandow S and Sch\"{u}tz G 1994 
\newblock On $U_q[SU(2)]$-symmetric driven diffusion 
\newblock {\it Europhys. Lett.} {\bf 26} 7

\bibitem{Schu93b} 
Sch\"utz G and Domany E 1993
\newblock Phase transitions in an exactly soluble one-dimensional
exclusion process
\newblock {\it J. Stat. Phys.} {\bf 72} 277

\bibitem{Derr93b}
Derrida B, Evans M R, Hakim V and Pasquier V  1993
\newblock Exact solution of a 1D asymmetric exclusion process using a matrix
formulation
\newblock {\it J. Phys. A} {\bf 26} 1493

\bibitem{Ligg75} Liggett T M 1975 
\newblock Ergodic theorems for the asymmetric simple exclusion process
\newblock {\it Trans. Amer. Math. Soc.} {\bf 213} 237

\bibitem{Schu94}  Sch\"{u}tz G and Sandow S  1994 
\newblock Non-abelian symmetries of stochastic processes: derivation of 
correlation functions for random vertex models and disordered interacting 
many-particle systems
\newblock {\it Phys. Rev. E} {\bf 49} 2726

\bibitem{Sepp99} Sepp\"al\"ainen T 1999 
\newblock Existence of hydrodynamics for the totally asymmetric simple 
$K$-exclusion process 
\newblock {\it Ann. Prob.} {\bf 27} 361

\bibitem{Chow00} Chowdhury D, Santen L and Schadschneider A 2000
\newblock Statistical Physics of Vehicular Traffic and some Related Systems
\newblock {\it Phys. Rep.} {\bf 329} 199

\bibitem{Helb01} Helbing D 2001 
\newblock Traffic and related self-driven many-particle systems 
\newblock {\it Rev. Mod. Phys.} {\bf 73} 1067

\bibitem{Spit70} Spitzer F 1970 
\newblock Interaction of Markov Processes
\newblock {\it Adv. Math.} {\bf 5} 246

\bibitem{Andj82} Andjel E D 1982
\newblock Invariant measures for the zero range process.
\newblock {\em Ann. Prob.} {\bf 10} 525

\bibitem{Redn97} Redner S 1997 
\newblock Scaling theories of diffusion-controlled and ballistically
controlled bimolecular reactions
\newblock in: {\it Nonequilibrium Statistical Mechanics in One Dimension} 
ed V Privman
\newblock (Cambridge: Cambridge University Press)

\bibitem{Alim00} Alimohammadi M and Ahmadi N 2000  
\newblock Class of integrable reaction-diffusion processes
\newblock {\it Phys. Rev. E} {\bf 62} 1674

\bibitem{Lahi97} Lahiri R and Ramaswamy S 1997
\newblock Are steadily moving crystals unstable?
\newblock {\it Phys. Rev. Lett.} {\bf 79} 1150

\bibitem{Popk01}  Popkov V and Peschel I 2001  
\newblock Symmetry breaking and phase coexistence in a driven diffusive
two-channel system
\newblock {\it Phys. Rev. E } {\bf 64} 026126

\bibitem{Toth02} T\'oth B and Valk\'o B 2002
\newblock Onsager relations and Eulerian hydrodynamics for systems
with several conservation laws
\newblock {\it math.PR/0210426}

\bibitem{Baxt82} Baxter R J 1982
\newblock {\it Exactly solved models in statistical mechanics}
\newblock (New York: Academic Press)

\bibitem{Kand90} Kandel D, Domany E and Nienhuis B 1990
\newblock A six-vertex model as a diffusion problem -- derivation of 
correlation functions
\newblock {\it J. Phys. A} {\bf 23} L755

\bibitem{Schu93c} Sch\"utz G 1993 
\newblock Time-dependent correlation functions in a one-dimensional 
asymmetric exclusion process
\newblock {\it Phys. Rev. E} {\bf 47} 4265

\bibitem{Hone96} Honecker A and Peschel I 1996 
\newblock Matrix product states for a one-dimensional lattice gas with
parallel dynamics
\newblock {\it J. Stat. Phys.} {\bf 88} 319

\bibitem{Pigo02} Pigorsch C and Sch\"utz G 2000
\newblock Shocks in the asymmetric simple exclusion process in a 
discrete-time update
\newblock {\it J. Phys. A} {\bf 33} 7919

\bibitem{Alca99a} Alcaraz F C and Bariev R Z 1999
\newblock Exact solution of a vertex model with an unlimited number of 
states per bond
\newblock {\it J. Phys. A} {\bf 232} L25

\bibitem{Alca94} Alcaraz F C, Droz M, Henkel M and Rittenberg V 1994
\newblock Reaction-diffusion processes, critical dynamics and quantum chains
\newblock {\it Ann. Phys. (NY)} {\bf 230} 250

\bibitem{Schu95a} Sch\"utz G M  1995
\newblock Reaction-diffusion processes of hard-core particles
\newblock {\it J. Stat. Phys.} {\bf 79} 243

\bibitem{Dahm95} Dahmen S R 1995 
\newblock Reaction-diffusion processes described by 3-state quantum chains 
and integrability
\newblock {\it J. Phys. A.} {\bf 28} 905

\bibitem{Fuji97} Fujii Y and Wadati M 1997
\newblock Reaction-diffusion processes with multi-species of particles
\newblock {\it J. Phys. Soc. Jap.} {\bf 66} 3770

\bibitem{Mobi01} Mobilia M and Bares P-A 2001
\newblock Soluble two-species diffusion-limited models in arbitrary 
dimensions
\newblock {\it Phys. Rev. E} {\bf 63} 036121

\bibitem{Krug91b} Krug J and Spohn H 1991 
\newblock Kinetic roughening of growing
surfaces, in: {\it Solids Far From Equilibrium} ed C Godr\'eche 
\newblock (Cambridge: Cambridge University Press)

\bibitem{Sasa98} Sasamoto T and Wadati M 1998
\newblock Exact results for one-dimensional totally asymmetric 
diffusion models
\newblock {\it J. Phys. A.} {\bf 31} 6057

\bibitem{Alca99} Alcaraz F C and Bariev R 1999
\newblock Exact Solution of the Asymmetric Exclusion Model 
with Particles of Arbitrary Size
\newblock {\it Phys. Rev. E} {\bf 60} 79

\bibitem{Ferr02} Ferreira A and Alcaraz F C 2002
\newblock Anomalous tag diffusion in the asymmetric exclusion 
model with particles of arbitrary sizes
\newblock {\it Phys. Rev. E} {\bf 65} 052102

\bibitem{Laka02} Lakatos G and Chou T 2003
\newblock Totally asymmetric
exclusion process with particles of arbitrary size
\newblock {\it J. Phys. A.} {\bf 36} 2027

\bibitem{Shaw03} Shaw L B, Zia R K P and Lee K H 2003
\newblock Modeling, simulations, and analyses of protein 
synthesis: Driven lattice gas with extended objects
\newblock {\it cond-mat/0302128}

\bibitem{Benj96} Benjamini I, Ferrari P A and Landim C 1996
\newblock Asymmetric conservative processes with random rates
\newblock {\it Stochastic Process. Appl.} {\bf 61} 181

\bibitem{Sepp99b} Sepp\"al\"ainen T and Krug J 1999 
\newblock Hydrodynamics and platoon formation for a
totally asymmetric exclusion model with particlewise disorder
\newblock {\it J. Stat. Phys.} {\bf 95} 525

\bibitem{Barm02} Barma M and Jain K 2002
\newblock Locating the minimum: Approach to equilibrium in a disordered,
  symmetric zero range process.
\newblock {\it Pramana - J.\ Phys.}, {\bf 58} 409

\bibitem{Evan96} Evans M R 1996 
\newblock Bose-Einstein condensation in
disordered exclusion models and relation to traffic flow
\newblock {\it Europhys. Lett.} {\bf 36} 13

\bibitem{Krug96} Krug J and Ferrari P A 1996
\newblock Phase Transitions in driven diffusive systems with random rates
\newblock {\it J. Phys. A} {\bf 29} L465

\bibitem{Kari99a} Karimipour V 1999
\newblock A multispecies asymmetric simple exclusion process and its
relation to traffic flow
\newblock {\it Phys. Rev. E} {\bf 59} 205

\bibitem{Kari99b} Karimipour V 1999
\newblock A multispecies asymmetric simple exclusion process, steady state and
correlation functions on a periodic lattice
\newblock {\it Europhys. Lett.} {\bf 47} 304

\bibitem{Khor00} Khorrami M and Karimipour V 2000
\newblock Exact determination of the phase structure of the p-species
asymmetric exclusion process
\newblock {\it J. Stat. Phys.} {\bf 100} 999

\bibitem{Beng99} Bengrine M, Benyoussef A Ez-Zahraouy H, Krug J,
Loulidi M and Mhirech F 1999
\newblock A simulation study of an asymmetric exclusion model with 
open boundaries and random rates 
\newblock {\it J. Phys. A} {\bf 32} 2527

\bibitem{Kolo98}  Kolomeisky A B, Sch\"{u}tz G M, Kolomeisky E B and
Straley J P 1998 
\newblock Phase diagram of one-dimensional driven lattice gases with open
boundaries
\newblock {\it J. Phys. A} {\bf 31} 6911

\bibitem{Niva94} Nivarthi S S, McCormick C V and Davis H T 1994
\newblock Diffusion anisotropy in molecular sieves - a Fourier transform 
PFG NMR-study of Methane in AlPO4-5
\newblock {\it Chem. Phys. Lett.} {\bf 229} 297

\bibitem{Kukl96} Kukla V, Kornatowski J, Demuth D, Girnus I,
Pfeifer H, Rees L V C, Schunk S, Unger K and K\"arger J 1996
\newblock NMR studies of single-file diffusion in unidimensional channel 
zeolites
\newblock {\it Science} {\bf 272} 702

\bibitem{Lei96} Lei G D, Carvill B T, and Sachtler W M H 1996
\newblock Single file diffusion in mordenite channels: Neopentane 
conversion and H/D exchange as catalytic probes
\newblock {\it Appl. Catal. A} {\bf 142} 347

\bibitem{Jobi97} Jobic H, Hahn K, K\"arger J, B\'{e}e M, Tuel A
Noack M, Girnus I and Kearley G J 1997
\newblock Unidirectional and single-file diffusion of molecules in 
one-dimensional channel systems. A quasi-elastic neutron scattering study
\newblock {\it J. Phys. Chem B} {\bf 101} 5834

\bibitem{Song00} Song L and Rees L V C 2000
\newblock Diffusion of propane in theta-1 and silicalite-1 zeolites
\newblock {\it Micropor. Mesopor. Mater.} {\bf 41} 193

\bibitem{Meer00} Meersmann T, Logan J W, Simonutti R, Caldarelli S,
Comotti A, Sozzani P, Kaiser L G and Pines A 2000
\newblock Exploring single-file diffusion in one-dimensional nanochannels 
by laser-polarized Xe-129 NMR spectroscopy
\newblock {\it J. Phys. Chem A} {\bf 104} 11665

\bibitem{Wei00} Wei Q-H, Bechinger C and Leiderer P 2000
\newblock Single-File diffusion of colloids in one-dimensional channels
\newblock {\it Science} {\bf 287} 625

\bibitem{Katz84} Katz S, Lebowitz J L and Spohn H 1984 
\newblock Nonequilibrium steady states of stochastic lattice gas 
models of fast ionic conductors
\newblock {\it J. Stat. Phys.} {\bf 34} 497

\bibitem{Sand95}  Sandow S, Trimper S and Mukamel D 1995
\newblock Asymmetric exclusion model for mixed ionic conductors
\newblock {\it Phys. Rev. B} {\bf 51} 2805

\bibitem{Chou99} Chou T and Lohse D 1999
\newblock Entropy-driven pumping in zeolites and biological channels
\newblock {\it Phys. Rev. Lett.} {\bf 82} 3552

\bibitem{Alex78a} Alexander S and Pincus P 1978
\newblock Diffusion of labeled particles on a one-dimensional chain
\newblock {\it Phys. Rev. B} {\bf 18} 2011

\bibitem{vanB83} van Beijeren H, Kehr K W and Kutner R 1983 
\newblock Diffusion in concentrated lattice gases. III. Tracer diffusion on a
one-dimensional lattice
\newblock {\it Phys. Rev. B}  {\bf 28}  5711

\bibitem{Arra83} Arratia R 1983,
\newblock The Motion of a Tagged Particle in the Simple Symmetric Exclusion
System in Z 
\newblock {\it Ann. Prob.} {\bf 11} 362

\bibitem{Karg92}
K\"arger J and Ruthven D M 1992 
\newblock {\em Diffusion in zeolites},
\newblock (John Wiley: New York)

\bibitem{Burl96} Burlatsky S F, Oshanin G, Moreau M and Reinhardt W P 1996
\newblock Motion of a driven tracer particle in a one-dimensional symmetric
lattice gas 
\newblock {\it Phys. Rev. E} {\bf 54}  3165

\bibitem{deMa85} De Masi A and Ferrari P A 1985 
\newblock Self-Diffusion in 
One-Dimensional Lattice Gases in the Presence of an External Field
\newblock J. Stat. Phys. {\bf 38} 603

\bibitem{vanB91} van Beijeren H 1991 
\newblock Fluctuations in the motion of mass
and of patterns in one-dimensional driven diffusive systems 
\newblock J. Stat. Phys. {\bf 63} 47

\bibitem{Maju91} Majumdar S N and Barma M 1991 
\newblock Tag diffusion in driven
systems, growing interfaces, and anomalous fluctuations 
\newblock {\it Phys. Rev. B} {\bf 44}  5306

\bibitem{Derr93a} Derrida B, Evans M R and Mukamel D 1993 
\newblock Exact diffusion
constant for one-dimensional asymmetric exclusion models 
\newblock {\it J. Phys. A} {\bf 26} 4911

\bibitem{Derr97a} Derrida B and Mallick K 1997 
\newblock Exact diffusion
constant for the one-dimensional partially asymmetric exclusion model
\newblock {\it J. Phys. A} {\bf 30} 1031

\bibitem{Gwa92} Gwa L-H and Spohn H 1992 
\newblock Bethe solution for the dynamical scaling exponent of the noisy 
Burgers equation 
\newblock {\it Phys. Rev. A} {\bf 46}  844

\bibitem{Kim95} Kim D 1995 
\newblock Bethe Ansatz solution for crossover scaling
functions of the asymmetric XXZ chain and the Kardar-Parisi-Zhang growth model
\newblock {\it Phys. Rev. E} {\bf 52}  3512


\bibitem{Ferr91} Ferrari P A, Kipnis C and Saada E 1991 
\newblock Microscopic structure of travelling waves in the asymmetric 
simple exclusion
\newblock {\it Ann. Prob.} {\bf 19} 226

\bibitem{Spoh91} Spohn H 1991
\newblock {\it Large Scale Dynamics of Interacting particles} 
\newblock (Berlin: Springer)

\bibitem{Ferr94a} Ferrari P A and Fontes L R G 1994 
\newblock Shock fluctuations in the asymmetric simple exclusion process 
\newblock {\it Probab. Theory Relat. Fields} {\bf 99} 305

\bibitem{Derr97} Derrida B, Lebowitz J L and Speer E R 1997 
\newblock Shock profiles for the Asymmetric Simple Exclusion Process in One 
Dimension
\newblock {\it J. Stat. Phys. 89} 135

\bibitem{Ferr00} Ferrari P A, Fontes L R G and Vares M E 2000
\newblock The Asymmetric Simple Exclusion Model with Multiple Shocks 
\newblock {\it Ann. Inst. H. Poincar\'e PR} {\bf 36} 109

\bibitem{Beli02} Belitsky V and Sch\"utz G M 2002
\newblock Diffusion and scattering of shocks in the partially 
asymmetric simple exclusion process
\newblock {\it El. J. Prob.} {\bf 7} paper 11 1-21

\bibitem{Ferr93} Ferrari P A and Kipnis C 1993
\newblock Second class particles in the rarefaction fan
\newblock {\it Ann. Inst. H. Poincar\'e PR} {\bf 31} 143

\bibitem{Chal02} Challet D, Willmann R D and Sch\"utz G M 2002
\newblock Exact Hurst exponent and crossover behavior in a limit order market 
model
\newblock {\it Physica A} {\bf 316} 430

\bibitem{Derr99} Derrida B and Evans M R 1999
\newblock Bethe ansatz solution for a defect particle in the
asymmetric exclusion process
\newblock {\it J. Phys. A} {\bf 32} 4833

\bibitem{Gree35} Greenshields B D 1935 
\newblock in: {\it Proceedings of the Highway Research Board} Vol 14 
\newblock (Washington: Highway Research Board)

\bibitem{Nage92} Nagel K and Schreckenberg M 1992 
\newblock A cellular automaton model for freeway traffic
\newblock {\it J. Phys. I} {\bf 2} 2221

\bibitem{Popk01b} Popkov V, Santen L, Schadschneider A and Sch\"utz G M 2001
\newblock Boundary-induced phase transition in traffic flow
\newblock {\it J. Phys. A} {\bf 34} L45

\bibitem{Mall96} Mallick K 1996 
\newblock Shocks in the asymmetry exclusion model with an impurity 
\newblock {\it J. Phys. A} {\bf 29} 5375

\bibitem{Lee97} Lee H-W, Popkov V and Kim D 1997 
\newblock Two-way traffic flow: Exactly solvable model of traffic jam 
\newblock {\it J. Phys. A} {\bf 30} 8497

\bibitem{Jafa00} Jafarpour F H 2000 
\newblock Exact solution of an exclusion model in the presence of a moving 
impurity on a ring 
\newblock {\it J. Phys. A} {\bf 33} 8673

\bibitem{MacD68} MacDonald J T, Gibbs J H and Pipkin A C 1968
\newblock Kinetics of biopolymerization on nucleic acid templates
\newblock {\it Biopolymers} {\bf 6} 1 

\bibitem{MacD69} MacDonald J T, Gibbs J H 1969
Concerning the kinetics of polypeptide synthesis on polyribosomes
\newblock {\it Biopolymers} {\bf 7} 707 (1969).

\bibitem{Heij87} von Heijne G, Blomberg C and Liljenstr\"om H 1987
\newblock Theoretical Modelling of protein synthesis
\newblock {\it J. theor. Biol.} {\bf 125} 1 

\bibitem{Schu97} Sch\"utz G M 1997
\newblock The Heisenberg chain as a dynamical model
for protein synthesis - Some theoretical and experimental results
\newblock {\it  Int. J. Mod. Phys. B} {\bf 11} 197

\bibitem{Lipo} Lipowsky R, Klumpp S and Nieuwenhuizen T M 2001
\newblock Random walks of cytoskeletal motors in open and closed compartments
\newblock  {\it Phys. Rev. Lett.}  {\bf 87} 108101

\bibitem{Kolo} Mirin  N and Kolomeisky AB 2003
\newblock Effect of Detachments in Asymmetric Simple Exclusion Processes 
\newblock  {\it J. Stat. Phys.}  {\bf 110} 811

\bibitem{Parm03} Parmeggiani A, Franosch T and Frey E 2003
\newblock Phase coexistence in driven one-dimensional transport
\newblock {\it Phys. Rev. Lett.} {\bf 90} 086601

\bibitem{Chow02} Chowdhury D, Guttal V, Nishinari K and Schadschneider A 2002
\newblock A cellular-automata model of flow in ant trails: non-monotic
variation of speed with density
\newblock {\it J. Phys. A} {\bf 35} L573

\bibitem{Burd02} Burd M, Archer D, Aranwela N and Stradling D J 2002
\newblock Traffic dynamics of the leaf-cutting ant, Atta cephalotes
\newblock {\it Am. Natur.} {\bf 159} 283

\bibitem{Oloa98} O'Loan O J, Evans M R and Cates M E 1998
\newblock Jamming transition in a homogeneous one-dimensional system:
The bus route model
\newblock {\it Phys. Rev. E} {\bf 58} 1404

\bibitem{MacL02} MacLeish T C B 2002
\newblock Tube theory of entangled polymer dynamics
\newblock {\it Adv. Phys.} {\bf 51} 1379

\bibitem{Doi86} Doi M and Edwards S F 1986
\newblock {\it The Theory of Polymer Dynamics} 
\newblock (Oxford: Oxford University Press)

\bibitem{deGe79} de Gennes P G 1979
\newblock {\it Scaling Concepts in Polymer Physics} 
\newblock (Ithaca: Cornell University Press)

\bibitem{Rubi87} Rubinstein M 1987 
\newblock Discretized model of entangled-polymer dynamics
\newblock {\it Phys. Rev. Lett.} {\bf 59} 1946

\bibitem{Perk94} Perkins T T, Smith D E and Chu S 1994
\newblock Direct observation of tube-like motion of a single polymer-chain
\newblock {\it Science} {\bf 264} 819

\bibitem{Schu99} Sch\"utz G M 1999
\newblock Non-equilibrium relaxation law for entangled polymers
\newblock {\it Europhys. Lett.} {\bf 48} 623

\bibitem{Duke89} Duke T A J 1989 
\newblock Tube model of field-inversion electrophoresis
\newblock {\it Phys. Rev. Lett.} {\bf 62} 2877

\bibitem{vanL92} van Leeuwen J M J and Kooiman A 1992
\newblock The drift velocity in the Rubinstein-Duke model for electrophoresis
\newblock {\it Physica A} {\bf 184}, 79

\bibitem{Prae96} Pr\"ahofer M and Spohn H 1996
\newblock Bounds on the diffusion constant for the Rubinstein-Duke model 
of electrophoresis
\newblock {\it Physica A} {\bf 233}, 191

\bibitem{Bark94} Barkema G T, Marko J F and Widom B 1994
\newblock Electrophoresis of charged polymers: Simulation and scaling in a 
lattice model of reptation
\newblock {\it Phys. Rev. E} {\bf 49} 5303

\bibitem{Bark96} Barkema G T, Caron C and Marko J F 1996
\newblock Scaling properties of gel electrophoresis of DNA
\newblock {\it Biopolymers} {\bf 38} 665

\bibitem{Aalb96} Aalberts D P and van Leeuwen J M J 1996
\newblock Dynamic symmetry breaking in a model of polymer reptation
\newblock {\it Electrophoresis} {\bf 17} 1003

\bibitem{Will02} Willmann R D 2002
\newblock Diffusion coefficient for reptation of polymers with kinematic 
disorder
\newblock {\it J. Chem. Phys.} {\bf 116} 2688

\bibitem{Carl01} Carlon E, Drzewinski A and van Leeuwen J M J 2001
\newblock Crossover behavior for long reptating polymers
\newblock {\it Phys. Rev. E} {\bf 64} 010801

\bibitem{Paes02} Paessens M and Sch\"utz G M 2002
\newblock DMRG studies of the effect of constraint release on the 
viscosity of polymer melts
\newblock {\it Phys. Rev. E} {\bf 66} 012806

\bibitem{Kawa66} Kawasaki K 1966 
\newblock Diffusion constants near critical point for time-dependent
Ising models. I.
\newblock {\it Phys. Rev.} {\bf 145} 224

\bibitem{Font03} Fontes L R G and Siqueira A 2003
\newblock in preparation

\bibitem{Meak86} Meakin P, Ramanlal P, Sander L M and Ball R C 1986
\newblock Ballistic deposition on surfaces 
\newblock {\it Phys. Rev. A} {\bf 34}  5091

\bibitem{Plis87} Plischke M, R\'acz Z and Liu D 1987
Time-reversal invariance and universality of two-dimensional growth models
\newblock {\it Phys. Rev. B} {\bf 35} 3485

\bibitem{Kard86} Kardar M, Parisi G and Zhang Y-C 1986 
\newblock Dynamic scaling of growing interfaces
\newblock {\it  Phys. Rev. Lett.} {\bf 56} 889

\bibitem{Arnd98a} Arndt P F, Heinzel T and Rittenberg V 1998
\newblock Stochastic models on a ring and quadratic algebras. The 
three-species diffusion problem
\newblock {\it J. Phys. A} {\bf 31} 833

\bibitem{Alca93} Alcaraz F C and Rittenberg V 1993
\newblock Reaction-diffusion processes as physical realizations of Hecke 
algebras
\newblock {\it Phys. Lett. B} {\bf 314} 377 

\bibitem{Popk02b} Popkov V, Fouladvand M E and Sch\"utz G M 2002
\newblock A sufficient criterion for integrability of stochastic
many-body dynamics and quantum spin chains
\newblock {\it J. Phys. A} {\bf 35} 7187

\bibitem{Derr98} Derrida B 1998
\newblock An exactly soluble non-equilibrium system: 
The asymmetric simple exclusion process
\newblock {\it Phys. Rep.} {\bf 301} 65

\bibitem{Stin95a} Stinchcombe R B and Sch\"utz G M 1995
\newblock Operator algebras for stochastic dynamics and the Heisenberg chain
\newblock {\it Europhys. Lett.} {\bf 29} 663

\bibitem{Stin95b} Stinchcombe R B and Sch\"utz G M 1995
\newblock Application of operator algebras to stochastic dynamics and the 
Heisenberg chain
\newblock {\it Phys. Rev. Lett.} {\bf 75} 140

\bibitem{Schu98b} Sch\"utz G M 1998
\newblock Dynamic Matrix Ansatz for Integrable 
Reaction-Diffusion Processes
\newblock {\it Eur. Phys. J. B} {\bf 5}, 589

\bibitem{Kreb97} Krebs K and Sandow S 1997
\newblock Matrix product eigenstates for one-dimensional stochastic models
and quantum spin chains
\newblock {\it J. Phys. A} {\bf 30} 3165

\bibitem{Hinr96c} Hinrichsen H, Sandow S and Peschel I 1996
\newblock On matrix product ground states for reaction-diffusion models
\newblock {\it J. Phys. A} {\bf 29} 2643

\bibitem{Essl96} Essler F H L and  Rittenberg V 1996
\newblock Representation of the quadratic algebra and partially asymmetric
diffusion with open boundaries
\newblock {\it J. Phys. A} {\bf 29} 3375

\bibitem{Mall97} Mallick K and Sandow S 1997
\newblock Finite-dimensional representations of the quadratic algebra: 
Applications to the exclusion process
\newblock {\it J. Phys. A} {\bf 30} 4513

\bibitem{Isae02} Isaev A P, Pyatov P N and Rittenberg V 2001
\newblock Diffusion algebras
\newblock {\it J. Phys. A} {\bf 34} 5815

\bibitem{Derr93c} Derrida B, Janowsky S A, Lebowitz J L and Speer E R 1993
\newblock Microscopic shock profiles: Exact solution of a nonequilibrium 
system
\newblock {\it Europhys. Lett.} {\bf 22} 651

\bibitem{Kreb03} Krebs K, Jafarpour F H and Sch\"utz G M 2003
\newblock in preparation

\bibitem{Bout02} Boutillier C, Francois P, Mallick K and Mallick S 2002
\newblock A matrix ansatz for the diffusion of an impurity in the 
asymmetric exclusion process
\newblock {\it J. Phys. A} {\bf 35} 9703

\bibitem{Schu93a} Sch\"{u}tz G 1993 
\newblock Generalized Bethe ansatz solution
of a one-dimensional asymmetric exclusion process on a ring with blockage
\newblock {\it J. Stat. Phys.} {\bf 71} 471

\bibitem{Arnd98b} Arndt P F, Heinzel T and Rittenberg V 1998 
\newblock Spontaneous breaking of translational invariance in
one-dimensional stationary states on a ring
\newblock {\it J. Phys. A }{\bf 31} L45 

\bibitem{Wolf90} Wolf D E and Tang L-H 1990
\newblock Inhomogeneous growth processes
\newblock {\it  Phys. Rev. Lett.} {\bf 65} 1591

\bibitem{Jano92}
Janowsky S A and Lebowitz J L 1992 
\newblock Finite-size effects and shock fluctuations in the
asymmetric simple exclusion process
\newblock {\it  Phys. Rev. A} {\bf 45} 618

\bibitem{Bran91} Brandstetter H 1991 
\newblock Diplom Thesis, University of Munich (unpublished).

\bibitem{Reza91}  Rezakhanlou F 1991
\newblock Hydrodynamic limit for attractive particle systems on $\bbbz^d$
\newblock {\it Comm. Math. Phys.} {\bf 140} 417

\bibitem{Burg74} Burgers J M 1974
\newblock {\it The nonlinear diffusion equation} 
\newblock (Boston: Riedel)

\bibitem{Baha98} Bahadoran C 1998
\newblock Hydrodynamical limit for spatially heterogeneous simple 
exclusion processes
\newblock {\it Probab Theory Rel} {\bf 110} 287

\bibitem{Schu96} Sch\"utz G M 1996
\newblock An exactly solvable lattice model for inhomogeneous interface growth
\newblock {\it J. Phys. I (Paris)} {\bf 6} 1405

\bibitem{Lax73} Lax P D 1973   
\newblock {\it Hyperbolic systems of conservation laws and the mathematical 
theory of shock waves} 
\newblock (Philadelphia: Society for Industrial and Applied Mathematics) 

\bibitem{Hage01} Hager J S, Krug J, Popkov V and Sch\"utz G M 2001
\newblock Minimal current phase and universal boundary layers in driven 
diffusive systems
\newblock {\it  Phys. Rev. E} {\bf 63} 056110

\bibitem{Yau91} Yau H T 1991
Relative entropy and hydrodynamics of Ginzburg-Landau models
\newblock {\it  Lett. Math. Phys.}  {\bf 22} 63

\bibitem{Gros03} Grosskinsky S and Spohn H 2003
\newblock to appear in Resenhas IME USP

\bibitem{Popk03} Popkov V and Sch\"utz G M 2003
\newblock Shocks and excitation dynamics in a driven diffusive two-channel
system
\newblock {\it cond-math/0211659} to appear in J. Stat. Phys.

\bibitem{Gray82} Gray L F 1982
\newblock The positive rates problem for attractive nearest neighbour
systems on $\bbbz$
\newblock {\it Z. Wahrsch. verw. Gebiete} {\bf 61} 389

\bibitem{Gacs86} Gacs P 1986
\newblock Reliable computation with cellular automata
\newblock {\it J. Comput. Syst. Sci.} {\bf 32} 15

\bibitem{Gacs01} Gacs P 2001
\newblock Reliable cellular automata with self-organization
\newblock {\it J. Stat. Phys.} {\bf 103} 45

\bibitem{Gray01} Gray L F 2001
\newblock A reader's guide to Gacs's ``Positive Rates'' paper
\newblock {\it J. Stat. Phys.} {\bf 103} 1

\bibitem{Evan02} Evans M R, Kafri Y, Levine E and Mukamel D 2002
\newblock Phase transition in a non-conserving driven diffusive system
\newblock {\it J. Phys. A} {\bf 35} L433

\bibitem{Popk03b} Popkov V, Willmann R D, Rakos A, Kolomeisky A B and
Sch\"utz G M 2003
\newblock Localization of shocks in driven diffusive systems without particle 
number conservation
\newblock {\it cond-mat/0302208}

\bibitem{Savi85} Savit R and Ziff R 1985
\newblock Morphology of a class of growth models
\newblock {\it Phys. Rev. Lett.} {\bf 55} 2515

\bibitem{Krug90} Krug J, Kertesz J and Wolf D E 1990
\newblock Growth shapes and directed percolation
\newblock {\it Europhys. Lett.} {\bf 12} 113

\bibitem{Alon96} Alon U, Evans M R, Hinrichsen H and Mukamel D 1996
\newblock Roughening transition in a one-dimensional growth process
\newblock {\it Phys. Rev. Lett.} {\bf 76} 2746

\bibitem{Alon98} Alon U, Evans M R, Hinrichsen H and Mukamel D 1998
\newblock Smooth phases, roughening transitions, and novel exponents in 
one-dimensional growth models
\newblock {\it Phys. Rev. E} {\bf 57} 4997

\bibitem{Hinr97} Hinrichsen H, Livi R, Mukamel D and Politi A 1997
\newblock Model for nonequilibrium wetting in two dimensions
\newblock {\it Phys. Rev. Lett.} {\bf 79} 2710

\bibitem{Kodu98} Koduvely H M and Dhar D 1998
\newblock A model of subdiffusive interface dynamics with a local conservation
of minimal height
\newblock {\it J. Stat. Phys.} {\bf 90} 57

\bibitem{Helb99} Helbing D Mukamel D and Sch\"utz G M 1999
\newblock Global phase diagram of a one-dimensional driven lattice gas
\newblock {\it Phys. Rev. Lett.} {\bf 82} 10

\bibitem{Schuunpub} Sch\"utz G M
\newblock unpublished

\bibitem{Evan98a} Evans M R,  Kafri Y, Koduvely H M and Mukamel D 1998
\newblock Phase separation in one-dimensional driven diffusive systems.
\newblock {\it Phys. Rev. Lett.} {\bf 80} 425

\bibitem{Lahi00} Lahiri R and Ramaswamy S 2000
\newblock Strong phase separation in a model of sedimenting lattices
\newblock {\it Phys. Rev. E} {\bf 61} 1648

\bibitem{Evan98b} Evans M R,  Kafri Y, Koduvely H M and Mukamel D 1998
\newblock Phase separation and coarsening in one-dimensional driven diffusive 
systems: Local dynamics leading to long-range Hamiltonians
\newblock {\it Phys. Rev. E} {\bf 58} 2764

\bibitem{Clin02} Clincy M and Evans M R 2002
\newblock Phase transition in the ABC model
\newblock {\it cond-mat/0209674} 

\bibitem{Arnd99} Arndt P F, Heinzel T and Rittenberg V 1998 
\newblock Spontaneous breaking of translational invariance spatial
condensation in stationary states on a ring. I. The neutral system
\newblock {\it J. Stat. Phys.} {\bf 97} 1

\bibitem{Arnd02} Arndt P F and Rittenberg V 2002
\newblock Spontaneous breaking of translational invariance spatial
condensation in stationary states on a ring. II. The charged system
and the two-component Burgers equations
\newblock {\it J. Stat. Phys.} {\bf 107} 989

\bibitem{Raje00} Rajewsky N, Sasamoto T and Speer E R 2000
\newblock Spatial condensation for an exclusion process on a ring
\newblock {\it Physica A} {\bf 279} 123

\bibitem{Kafr02a} Kafri Y, Levine E, Mukamel D and T\"or\"ok J 2002
\newblock Sharp crossover and anomalously large correlation length in driven
systems
\newblock {\it J. Phys. A} {\bf 35} L459

\bibitem{Kafr02b} Kafri Y, Levine E, Mukamel D, Sch\"utz G M and
Willmann R D 2002
\newblock Novel phase separation transition in one-dimensional driven models
\newblock {\it cond-mat/0211269}

\bibitem{Sasa01} Sasamoto T and Zagier D 2001
\newblock On the existence of a phase transition for an exclusion process 
on a ring
\newblock {\it J. Phys. A} {\bf 34} 5033

\bibitem{Korn99} Korniss G, Schmittmann B and Zia R K P 1999
\newblock Long-range order in a quasi one-dimensional non-equilibrium 
three-state lattice gas
\newblock {\it Europhys. Lett.} {\bf 45} 431

\bibitem{Mett02} Mettetal J T, Schmittmann B and Zia R K P 2002
\newblock Coarsening dynamics of a quasi-one-dimensional driven lattice gas
\newblock {\it Europhys. Lett.} {\bf 58} 653

\bibitem{Godr} Godr\`eche C and Sandow S
\newblock unpublished

\bibitem{Nagy02} Nagy Z, Appert C and Santen L 2002
\newblock Relaxation times in the ASEP model using a DMRG method
\newblock {\it J. Stat. Phys.} {\bf 109} 623

\bibitem{Kafr02} Kafri Y, Levine E, Mukamel D, Sch\"utz G M and 
T\"or\"ok J 2002
\newblock Criterion for phase separation in one-dimensional driven systems.
\newblock {\it Phys. Rev. Lett.} {\bf 89} 035702

\bibitem{Blyt00} Blythe R A, Evans M R, Colaiori F and Essler F H L 2000
\newblock Exact solution of a partially asymmetric exclusion model 
using a deformed oscillator algebra
\newblock {\it J. Phys. A} {\bf 33} 2313

\bibitem{Krug90b} Krug J and Meakin P 1990
\newblock Universal finite-size effects in the rate of growth processes
\newblock {\it J. Phys. A} {\bf 23} L987

\bibitem{Krug94} Krug J and Tang L-H 1994
\newblock Disorder-induced unbinding in confined geometries
\newblock {\it Phys. Rev. E} {\bf 50} 104

\bibitem{Kafr00} Kafri Y, Biron D, Evans M R and Mukamel D 2000
Slow coarsening in a class of driven systems
\newblock {\it Eur. Phys. J. B}  {\bf 16} 669

\bibitem{Gros03b} Grosskinsky S, Sch\"utz G M and Spohn H
\newblock Condensation in the zero range process: stationary and dynamical 
properties
\newblock {\it cond-mat/0302079} 

\bibitem{Corn91} Cornell S J, Kaski K and Stinchcombe R B 1991
\newblock Domain scaling and glassy dynamics in a one-dimensional
Kawasaki-Ising model
\newblock {\it Phys. Rev. B} {\bf 44} 12263

\bibitem{Maju94} Majumdar S N, Huse D A, Lubachevsky B D 1994
\newblock Growth of long range correlations after a quench in 
conserved-order-parameter systems
\newblock {\it Phys. Rev. Lett.} {\bf 73} 182

\bibitem{Corn96} Cornell S J and Bray A 1996
\newblock Domain growth in a one-dimensional driven diffusive system
\newblock {\it Phys. Rev. E} {\bf 54} 1153

\bibitem{Spir99} Spirin V, Krapivsky P L and Redner S 1999
\newblock Coarsening in a driven Ising chain with conserved dynamics
\newblock {\it Phys. Rev. E} {\bf 60} 2670

\bibitem{Godr03} Godr\`eche C
\newblock Dynamics of condensation in zero-range processes
\newblock {\it cond-mat/0301156} 

\bibitem{Taka93} Takayasu M and Takayasu H 1993
\newblock {\it  Fractals} {\bf 1} 860

\bibitem{Scha97} Schadschneider A and Schreckenberg M 1997
\newblock Traffic flow models with 'slow-to-start' rules
\newblock {\it  Ann. Phys.} {\bf 6} 541

\bibitem{Benj96b} Benjamin S C, Johnson N F and Hui P M 1996
\newblock Cellular automata models of traffic flow along a highway 
containing a junction
\newblock {\it J. Phys. A} {\bf 29} 3119

\bibitem{Barl98} Barlovic R, Santen L, Schadschneider A and 
Schreckenberg M 1998
\newblock Metastable states in cellular automata for traffic flow
\newblock {\it Eur. Phys. J. B}  {\bf 5} 793

\bibitem{Nama02} Namazi A, Eissfeldt N, Wagner P and Schadschneider A 2002
\newblock Boundary-induced phase transitions in a space-continuous traffic
model with a non-unique flow-density relation
\newblock {\it Eur. Phys. J. B}  {\bf 30} 559

\bibitem{Evan95}  Evans M R, Foster D P, Godr\`eche C and Mukamel D 1995 
\newblock Asymmetric exclusion model with two species: spontaneous
symmetry breaking
\newblock {\it J. Stat. Phys.} {\bf 80}  69 

\bibitem{Joha00} Johansson K 2000
\newblock Shape fluctuations and random matrices
\newblock {\it Comm. Math. Phys.} {\bf 209} 437


\end{thebibliography}
\end{document}